\documentclass[%
 aip,
 amsmath,amssymb,
preprint,%
floatfix, 
]{revtex4-1}

\usepackage{graphicx}
\usepackage{placeins}
\usepackage{dcolumn}
\usepackage{bm}
\usepackage[USenglish]{babel}
\usepackage[utf8]{inputenc}
\usepackage[T1]{fontenc}
\usepackage{mathptmx}
\usepackage{etoolbox}
\usepackage{tcolorbox} 
\usepackage{tikz} 
\usepackage{pgfplots} 
\usepackage[hidelinks]{hyperref} 
\pgfplotsset{width=\linewidth, compat=1.6} 

\usepackage[normalem]{ulem} 
\usepackage{url}

\usepackage[left=0.5in, right=0.5in]{geometry} 

\makeatletter
\def\@email#1#2{%
 \endgroup
 \patchcmd{\titleblock@produce}
  {\frontmatter@RRAPformat}
  {\frontmatter@RRAPformat{\produce@RRAP{*#1\href{mailto:#2}{#2}}}\frontmatter@RRAPformat}
  {}{}
}%
\makeatother

\begin{document}
\newcommand{\removeindent}{6pt} 
\newcommand{\thistablearraystretch}{0.6}
\preprint{Preprint}

\title[Liquid distribution after head-on separation of two colliding immiscible liquid droplets]{Liquid distribution after head-on separation of two colliding immiscible liquid droplets}

\author{Johanna Potyka}
\affiliation{Institute of Aerospace Thermodynamics, University of Stuttgart\\ Pfaffenwaldring 31, 70569 Stuttgart, Germany}%

\author{Kathrin Schulte$^*$}%
 \email{kathrin.schulte@itlr.uni-stuttgart.de}
\affiliation{Institute of Aerospace Thermodynamics, University of Stuttgart\\ Pfaffenwaldring 31, 70569 Stuttgart, Germany}%

\author{Carole Planchette}
\affiliation{Institute of Fluid Mechanics and Heat Transfer, Technical University of Graz\\ Inffeldgasse 25 F, 8010 Graz, Austria}%

\date{\today}

\begin{abstract}
Head-on collisions of two immiscible liquid droplets lead to a collision complex, which may either remain stable in the form of a single compound drop, or fragment into two main daughter droplets. %
This paper investigates the liquid distribution developing in the two daughter droplets and which can be of three types. %
Either two encapsulated droplets (single reflex {separation}) form, or a single encapsulated drop plus a droplet made solely of the encapsulating liquid, which can be found either on the impact side (reflexive {separation}) or opposite to it (crossing {separation}). %
A large number of experimental and {simulation} data covering collisions with partial and total wetting conditions and with Weber and Reynolds numbers in the ranges of 2 - 720 and 66 - 1100, respectively, is analyzed. %
The conditions leading to the three mentioned liquid distributions are identified and described based on the decomposition of the collision in two phases: %
(i) radial extension of the compound droplet into a lamella and (ii) its relaxation into an elongated cylindrical droplet. %
In accordance with these two phases, two dimensionless parameters, $\Lambda = {\rho_i/\rho_o} {{We_i}^{-1/2}}$ and $N = {\nu_o/\nu_i}~{\sigma_{o}/\sigma_{io}}$, are derived, which are built on the collision parameters and liquid properties of the encapsulated inner droplet (i) and the outer droplet (o) only. %
In agreement with the proposed interpretation, the combination of these two parameters predicts the type of liquid distribution. %
The predictions are found to be in very good agreement with both experimental and numerical results.
\end{abstract}

\maketitle

\section{Introduction} \label{Sec:Introduction}
\begin{figure*}[tb!]
\null\hfill
\begin{minipage}[t]{0.29 \linewidth}
\includegraphics[width=\linewidth]{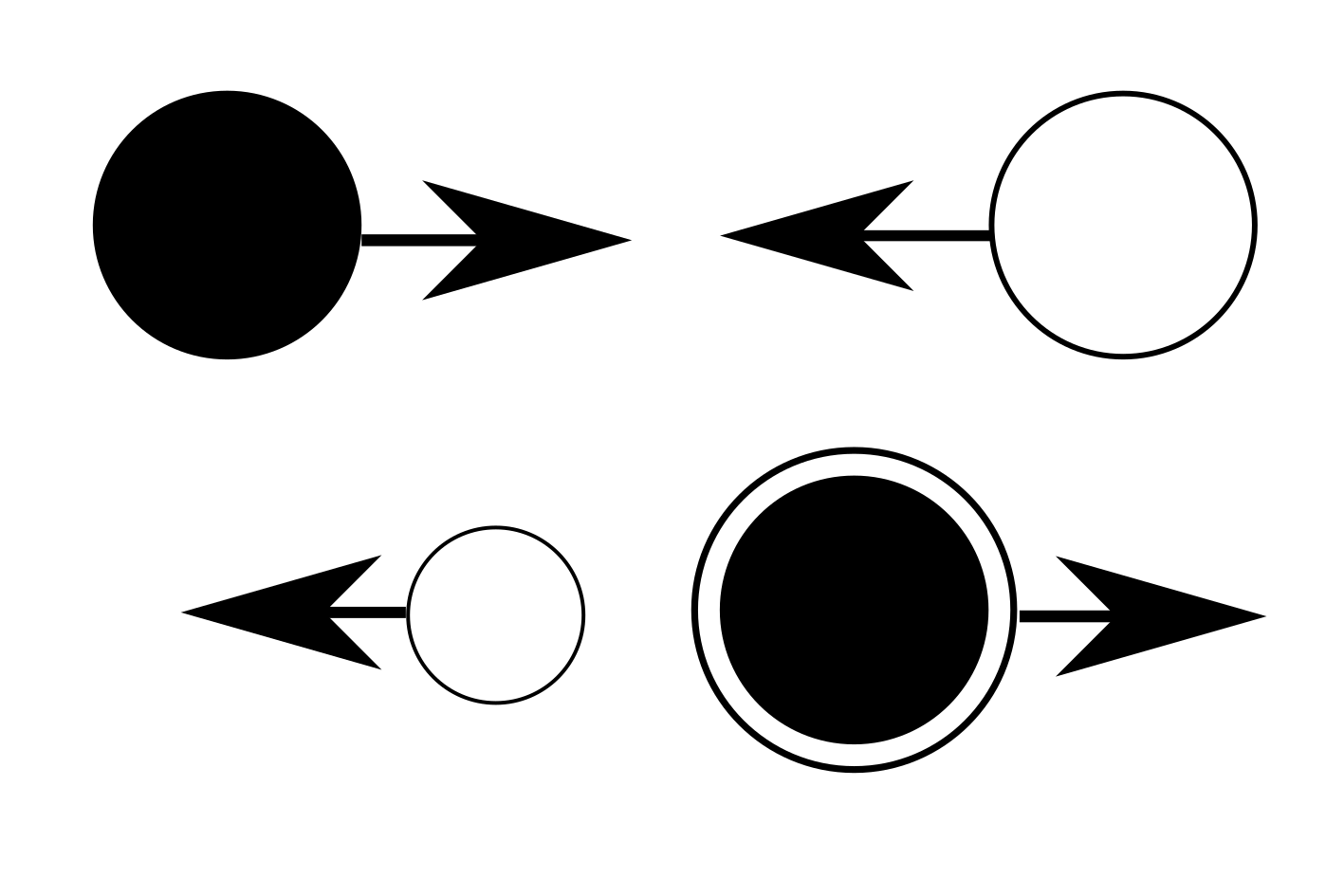}
\text{Crossing separation}
\end{minipage}
\hfill
\begin{minipage}[t]{0.29 \linewidth}
\includegraphics[width=\linewidth]{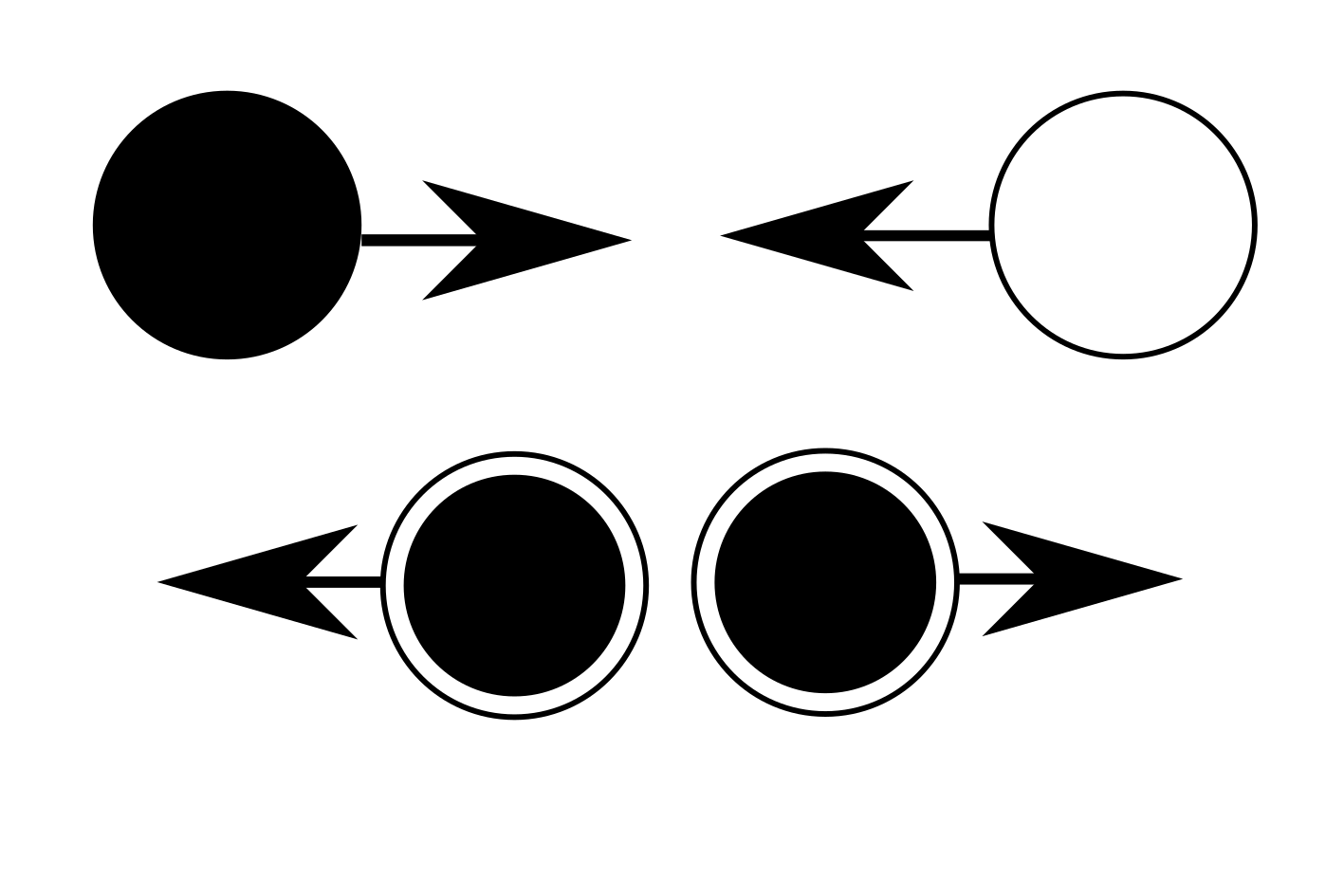}
\text{Single reflex separation}
\end{minipage}
\hfill
\begin{minipage}[t]{0.29 \linewidth}
\includegraphics[width=\linewidth]{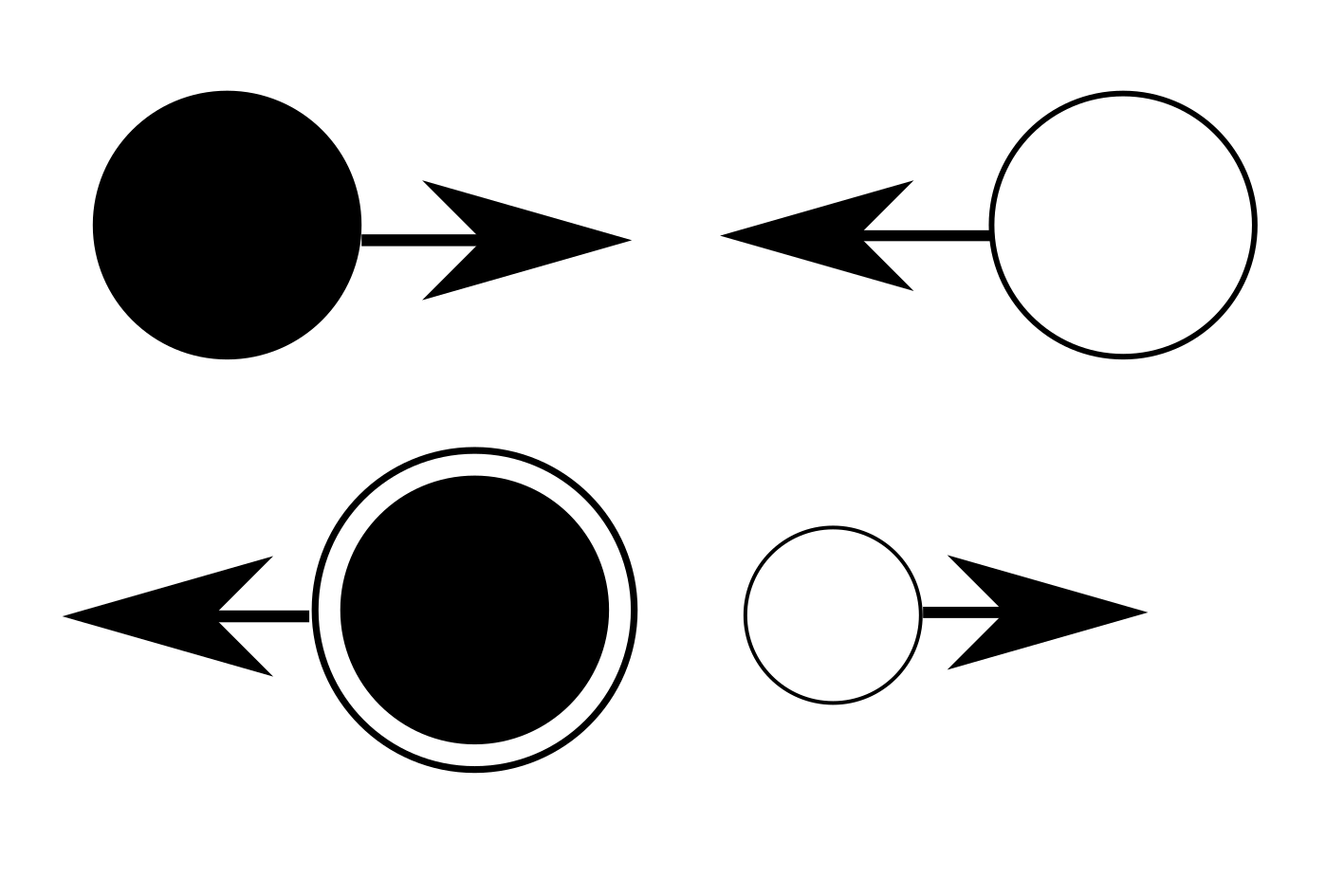}
\text{Reflexive separation}
\end{minipage}
\hfill\null
\caption{Head-on collisions of immiscible liquid drops leading to separation. Despite similar initial configurations (upper line), the final liquid distribution (lower line) varies, with  from left to right: crossing separation, single reflex separation and reflexive separation. Black/White color indicates the inner/outer or encapsulated/encapsulting liquid.}\label{Fig:OutcomeSketch}
\end{figure*}

Collisions of droplets of different and possibly immiscible liquids are a fundamental process in various technical and life-science applications. %
For example, in-air microfluidics as a chip-free platform technology uses a combination of droplets and jets of different liquids that collide in air to produce emulsions and modular 3D biomaterials such as hydrogel fibers for tissue engineering \cite{Visser2018, Kamperman2018, Marangon2023}. %
Binary collisions of partly or totally immiscible droplets is successfully used for encapsulating actives for food, cosmetic and pharmaceutical applications \cite{Yeo2004, Planchette2010}.

The possible outcomes of such collisions are, at first glance, very similar to those obtained when a single liquid is used.
These include (i) bouncing of the two droplets, (ii) merging of the two droplets, and (iii) separation, the latter occurring for head-on as well as off-center collisions \cite{Planchette2010, Chen2006a}. %
Thus, it may be tempting to generalize results obtained for binary collisions of droplets of the same liquids, which have been extensively studied experimentally \cite{Ashgriz1990,Jiang1992,Qian1997,Orme1997,Pan2019,Gotaas2007,Willis2000,Willis2003}, numerically \cite{Pan2005,Nobari1996,Sakakibara2008,Chen2020} and analytically \cite{Ashgriz1990,Jiang1992,Munnannur2007,Finotello2017}, to the case where immiscible liquids are used. %
Yet, while this analogy proved its robustness for describing the process phenomenology \cite{Planchette2017}, the existence of two liquid phases and an additional interface, which significantly modifies the flow fields \cite{Zhang2020}, brings this analysis to its limits. %
Important differences can for example be evidenced by comparing the magnitude of the viscous losses developing during the first instants of head-on collisions representing more than 90\% of the initial kinetic energy for immiscible liquids against approximately 60\% for drops of the same liquid \cite{Planchette2017}. %
Such characteristic differences call for detailed specific studies that have been carried out only to a limited extent so far. %
Chen and Chen first preformed dedicated experiments combining diesel and water droplets \cite{Chen2006a}. %
As a result, they identified the regime boundaries for bouncing, coalescence, head-on and off-center separations. %
The spatial distribution of the liquid of each droplet, here water and diesel, was found to be different than for the classical results obtained by Ashgriz and Poo using two droplets of water \cite{Ashgriz1990}. For off-center separation, i.e., when the collision eccentricity is important, the stretched ligament was found to be mostly made of diesel, the liquid with the smallest surface tension. %
For head-on separation, which refers to collisions with low eccentricity, the liquid distribution appeared to be more complex. %
The two liquids were found not to distribute symmetrically with respect to the collision point, which led Chen and Chen to re-name this regime single reflex in place of reflexive separation.%
No further interpretation of these observations was proposed, which may be explained by the lack of interfacial tension measurement, the variability of diesel composition and its low miscibility with water.

Some years after, \citet{Planchette2010, planchette_2012} conducted  more experiments  with totally immiscible liquids, namely aqueous glycerol solutions and silicon oils of various viscosity. %
They proposed to make use of the asymmetric liquid distribution observed during stretching separation to tune the shell thickness of the encapsulating liquid. %
The experiments demonstrated that the shell thickness was at first order fixed by the collision eccentricity, other parameters such as relative velocity and liquid properties having only second order effects. %
In case of head-on collisions, the authors noted that separation, which typically produces two main daughter droplets, could lead to different types of liquid distributions within these two daughter droplets. %
Three different types of liquid distributions were identified and named as crossing, single reflex and reflexive separation \cite{Planchette2010} depending on the location of the encapsulated (white) or inner liquid (black), cf.\ Fig.~\ref{Fig:OutcomeSketch}. %
Yet, no interpretation was provided and while efforts were put into the modeling of the separation threshold \cite{planchette_2012}, no analysis was proposed to predict the resulting type of liquid distribution.

At the same time, a theoretical model for head-on droplet spreading was deduced based on the investigations by \citet{Roisman2012}. %
Yet, the model focuses on the first phase of the collisions, i.e.,\ until the lamella reaches its maximal extension, but does not study the recoil phase and therefore does not provide information about the liquid distribution found after separation. %

Beyond these studies, several numerical works were performed providing a framework for the simulation of collisions of immiscible droplets \cite{Shi2016, HaghaniHassanAbadi2018, Woehrwag2018,Li2015}. %
\citet{Zhang2020} used numerical simulations for detailed investigations on the influence of interfacial tension on the initial evolution of the collision complex. %
However, they did not focus on the collision outcome. \\
To date, none of these studies propose a rational way to answer the question of where the two immiscible liquids are after a head-on collision, which is nevertheless essential for practical application. %

As already mentioned, for head-on collisions, i.e., for collisions with an eccentricity close to zero, two main droplets are produced just above the separation threshold. %
Their typical composition can significantly vary, with the presence of the encapsulated liquid in only one of the droplets - at the impact side or opposite to it - or in both drops. %
The delimitation of these three possible  distributions is the focus of the present study and is addressed here via the establishment of two-dimensionless parameters. Note that for relative velocities well above the separation threshold, a long ligament forms, which breaks-up in more than two droplets. %
Such outcomes  go beyond the scope of  the present study. %
The modeling approach we propose in Sec.~\ref{Sec:Discussion} is based on experimental results (Sec.~\ref{Sec:Experiments}) and further insights into the fluid distribution from numerical simulations (Sec.~\ref{Sec:Simulations}). %
Consideration about the shape taken by the deformed droplets during the first instants of the collision are used to distinguish the crossing regime from the two other regimes, while the distinction between single reflex and reflexive separation can be established by comparing their respective shape at longer timescale. %
The paper ends with the conclusions (Sec.~\ref{Sec:Conclusion}). Note that the data of all presented head-on collisions is provided as a dataset on the Data Repository of the University of Stuttgart \citep{DarusData2023}.

\section{Methods} \label{Sec:Methods}
This section first presents the parameters and quantities required to fully describe the collision between immiscible liquid droplets. In order to investigate the outcomes of such collisions, complementing experimental and numerical methods are used. The methods and the type of data they respectively provide are explained thereafter. The section ends with a list of the varied parameters and considered ranges for each  method. 

\subsection{Problem description}
\begin{figure}[!htb]
    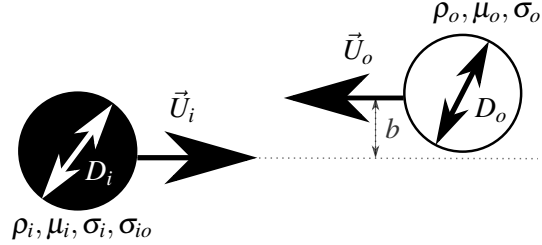
    \caption{Relevant parameters and quantities for describing the collisions of two drops made of immiscible liquids. %
    The relative velocity of the two droplets is $U_{rel}=|\vec{U}_i-\vec{U}_o|$.}
    \label{Fig:InputParameters}
\end{figure}
To describe the collisions of two immiscible liquid droplets, several quantities must be considered. %
First of all, the liquid properties of each liquid must be accounted for. %
As the liquids are immiscible, the droplet with the lower surface tension naturally spreads - partially at least - around the second droplet. %
Here and in the following, the former is indicated by the subscript "o" (for outer) in contrast to the latter indicated by the subscript "i" (for inner). %
Please note that as shown in the Figs.~\ref{Fig:OutcomeSketch}-\ref{Fig:InputParameters} and throughout the entire manuscript, the outer liquid appears white or transparent grey and the inner one is dark, both in the experimental and the simulations' visualizations. %
The relevant liquids' properties are the density, $\rho$, the viscosity, $\mu$, and the surface tension with respect to the ambient air, $\sigma$. %
The interfacial tension associated with the interface between the two immiscible liquids is denoted as $\sigma_{io}$. %
The wetting behaviour of the liquid combination is given by its spreading parameter $S=\sigma_i-(\sigma_o+\sigma_{io})$, which is positive/negative for total/partial wetting. %
In addition to these material properties, kinetic and geometric parameters must be considered. %
The relative velocity of the two droplets is defined as the difference between the inner and outer liquid droplet velocities providing $\vec{U}_{rel}=\vec{U_{i}}-\vec{U_{o}}$. %
The droplet sizes are given by their respective diameter $D_i$ and $D_o$. %
Finally, the eccentricity of the collision is quantified by  $b$, the distance separating their center of mass after projection perpendicularly to the relative velocity axis, cf.\ Fig.~\ref{Fig:InputParameters} for completeness. %
In practice, the eccentricity is normalized by the average diameter of the droplets and called the normalized impact parameter $X=b/(0.5 (D_i+D_o))$. %
In the present study, only head-on or quasi head-on collisions are considered ($X=0$ or $X<0.1$). %
We focus our investigation of the liquids' distribution on velocities corresponding to the separation threshold. %
Furthermore, the investigations are limited to equally-sized or quasi equally-sized droplets ($D_i=D_o$ or $ 2(D_i-D_o)/(D_i+D_o)< 5\%$). %
Thus in the following, we use $D$ without indices to designate both droplet diameters equally.

\begin{table}[!tb]
\renewcommand*{\arraystretch}{\thistablearraystretch}
\newcommand{\minus}{\scalebox{0.7}[1.0]{$-$}}
\newcommand{\doanewlineornot}{\\} %
    \centering
    \begin{tabular}{l|rcl rcl l}
    \hline
    \hline
        \doanewlineornot
        \bf{Input Data} & \multicolumn{3}{c}{\bf{Range Sim.}} & \multicolumn{3}{c}{\bf{Range Exp.}} & \bf{Unit}\\
        \doanewlineornot
            \hline
         \doanewlineornot
         $D_i$        & $100$ & -- & $800$  & $190$ & -- & $200$ & $\mathrm{\mu m}$ \\
         $D_o$        & $100$ & -- & $800$  & $190$ & -- & $210$ & $\mathrm{\mu m}$ \\
         \doanewlineornot         
        ${D_o / D_i}$  & $0.974$ & -- & $1.017$ & \multicolumn{3}{c}{$\approx 1.0$}    & -- \\
         \doanewlineornot
         $\rho_i$ & $500$ & --  & $2000$   & $1048$ & -- & $1154$  & $\mathrm{kg~m^{-3}}$ \\
         $\rho_o$ & $500$ & --  & $2000$   & $892$ & -- & $1934.9$ & $\mathrm{kg~m^{-3}}$ \\
         $\rho_g$ & \multicolumn{3}{c}{$1.19$}       & & -- &            & $\mathrm{kg~m^{-3}}$ \\
         \doanewlineornot         
         ${\rho_o / \rho_i}$ & $0.5$ & -- & $2.0$ & $0.78$ & -- & $1.72$ & -- \\
         \doanewlineornot
         $\mu_i$ & $1.0$ & -- & $8.0$     & $1.76$ & -- & $11.0$ & $\mathrm{mPas}$ \\
         $\mu_o$ & $1.5$ & -- & $7.0$     & $2.80$ & -- & $19.0$  & $\mathrm{mPas}$ \\
         $\mu_g$ & \multicolumn{3}{c}{$0.01824$} &  & -- &         & $\mathrm{mPas}$ \\
         \doanewlineornot         
         ${\mu_o / \mu_i}$ & $0.35$ & -- & $3.20$ & $0.35$ & -- &  $5.34$ & -- \\
         \doanewlineornot
         $\sigma_i$    & $35.0$ & -- & $83.1$  & $68.1$ & -- & $70.7$ & $\mathrm{mN~m^{-1}}$ \\ 
         $\sigma_o$    & $5.0$ & -- & $44.4$   & $17.8$ & -- & $20.7$ & $\mathrm{mN~m^{-1}}$ \\
         $\sigma_{io}$ & $5.7$ & -- & $54.8$ & $34.0$ & -- & $38.0$   & $\mathrm{mN~m^{-1}}$ \\
         \doanewlineornot         
         $ S $         & $\minus5.7$ & -- & $43.6$ & $12.9$ & -- & $15.7$ & $\mathrm{mN~m^{-1}}$ \\
         ${\sigma_o / \sigma_i}$ & $0.07$ & -- & $0.71$ &  $0.26$ & -- & $0.30$ & -- \\
         ${\sigma_{io} / \sigma_{i}}$ & $0.14$  & -- & $0.99$ & $0.49$ & -- &  $0.54$ & -- \\
         \doanewlineornot
         $U_{\mathrm{rel}}$ & $0.5$ & -- & $12.0$ & $2.32$ & -- & $6.96$ & $\mathrm{m~s^{-1}}$\\
         \doanewlineornot
         $We_i$ & $2.6$ & -- & $192$ & $16$ & -- & $159$ & -- \\
         $We_o$ & $6.7$ & -- & $720$ & $49$ & -- & $446$ & -- \\
         $Re_i$ & $67$ & -- & $1118$ & $68$ & -- & $452.5$ & -- \\
         $Re_o$ & $66$ & -- & $514$  & $64$ & -- & $240$ & -- \\
         \doanewlineornot
         \hline 
         \hline
    \end{tabular}
    \caption{Simulation and experimental setup data ranges. A detailed list of all cases investigated can be found in a dataset on the Data Repository of the University of Stuttgart \citep{DarusData2023}.}
    \label{tab:SimExpSetup}    
\end{table}
\begin{figure*}[!tb]
\centering
    \includegraphics[page=1,scale=1,trim=0cm 0.3cm 0cm 0cm,clip]{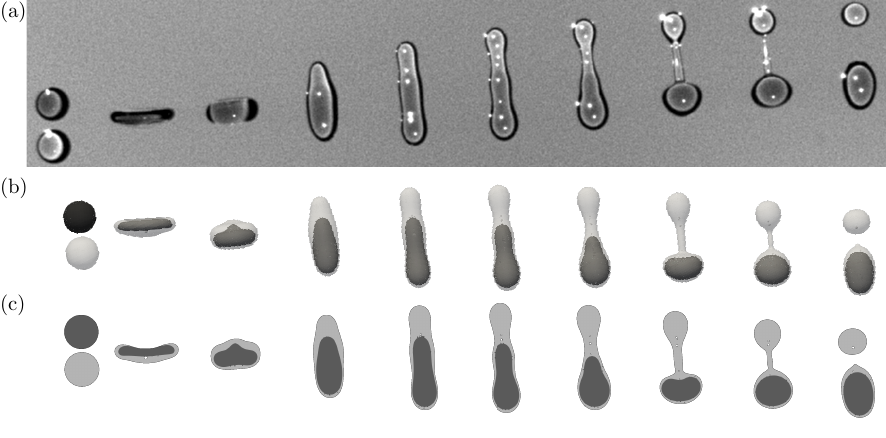}
    \caption{Crossing Separation: Comparison of (a) experiment and (b) outer view from simulations and (c) a section at the symmetry plane of the simulation. Collision of G50 inner (dark grey) and SOM5 outer liquid (transparent grey): droplet's diameter \mbox{$D_i = 192~\mathrm{\mu m}$}, droplet's diameter \mbox{$D_o = 194~\mathrm{\mu m}$}, relative velocity \mbox{$U_\mathrm{rel}= 3.74~\mathrm{m~s^{-1}}$}, impact parameter \mbox{$X = 0.0152$}, viscosity \mbox{$\mu_i= 5.55~\mathrm{mPas}$}, viscosity \mbox{$\mu_o = 5.1~\mathrm{mPas}$}, density \mbox{$\rho_i=1122.7~\mathrm{kg~m^{-3}}$}, density \mbox{$\rho_o=917~\mathrm{kg~m^{-3}}$}, surface tension \mbox{$\sigma_i= 68.3~\mathrm{mN~m^{-1}}$}, surface tension \mbox{$\sigma_o= 19.5~\mathrm{mN~m^{-1}}$}, interfacial tension \mbox{$\sigma_{io}= 35~\mathrm{mN~m^{-1}}$}. Simulation only: viscosity of air \mbox{$\mu_\mathrm{air} = 0.01824~\mathrm{mPas}$}, density of air \mbox{$\rho_\mathrm{air} = 1.19~\mathrm{kg~m^{-3}}$}.}
\label{Fig:ValidationPicturesCrossing}
\end{figure*}    
\begin{figure*} [!tb]
\centering
    \includegraphics[page=2,scale=1,trim=0cm 0cm 0cm 0cm, clip]{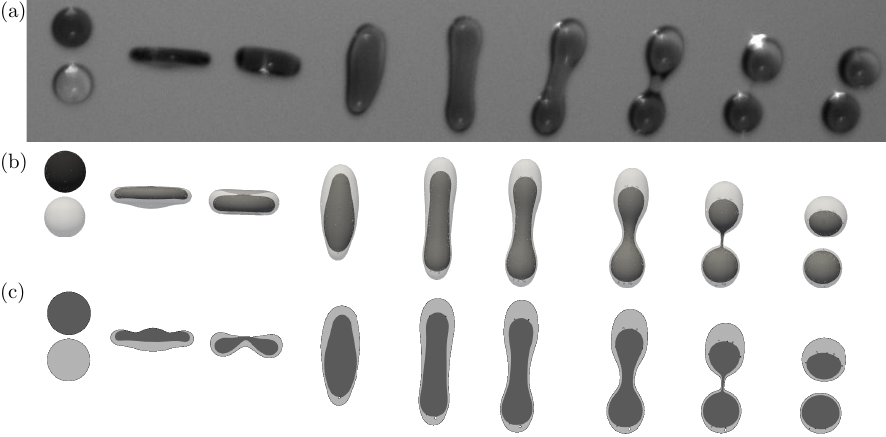} 
 \caption{Single reflex separation:  Comparison of (a) experiment and (b) outer view from simulations and (c) a section at the symmetry plane of the simulation. Collision of G50 inner (dark grey) and SOM10 outer liquid (transparent grey): droplet's diameter \mbox{$D_i = 200~\mathrm{\mu m}$}, droplet's diameter \mbox{$D_o = 197~\mathrm{\mu m}$}, relative velocity \mbox{$U_\mathrm{rel}= 4.56~\mathrm{m~s^{-1}}$}, impact parameter \mbox{$X = 0.009$}, viscosity \mbox{$\mu_i= 6~\mathrm{mPas}$}, viscosity \mbox{$\mu_o = 9.37~\mathrm{mPas}$}, density \mbox{$\rho_i=1126~\mathrm{kg~m^{-3}}$}, density \mbox{$\rho_o=937.2~\mathrm{kg~m^{-3}}$}, surface tension \mbox{$\sigma_i= 68.6~\mathrm{mN~m^{-1}}$}, surface tension \mbox{$\sigma_o= 20.1~\mathrm{mN~m^{-1}}$}, interfacial tension \mbox{$\sigma_{io}= 34.5~\mathrm{mN~m^{-1}}$}. Simulation only: viscosity of air \mbox{$\mu_\mathrm{air} = 0.01824~\mathrm{mPas}$}, density of air \mbox{$\rho_\mathrm{air} = 1.19~\mathrm{kg~m^{-3}}$}.}
\label{Fig:ValidationPicturesSingleReflex}
\end{figure*}    
\begin{figure*} [!tb]
\centering
    \includegraphics[page=4,scale=1,trim=0cm 0.8cm 0cm 0cm, clip]{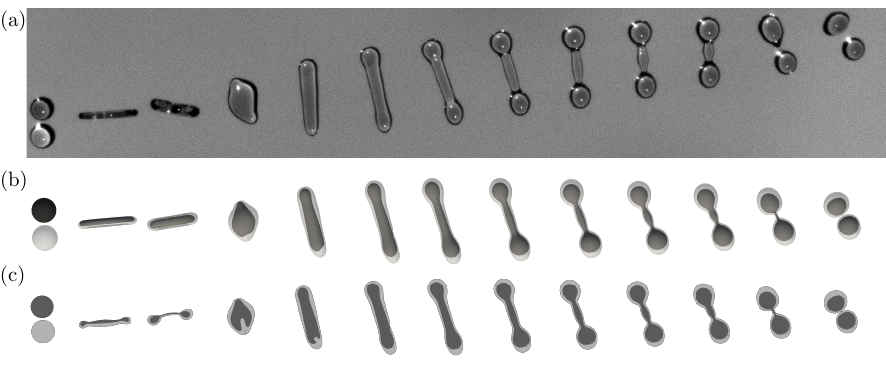} 
 \caption{Single reflex separation:  Comparison of (a) experiment and (b) outer view from simulations and (c) a section at the symmetry plane of the simulation. Collision of G50 inner (dark grey) and SOM20 outer liquid (transparent grey): droplet's diameter \mbox{$D_i = 202~\mathrm{\mu m}$}, droplet's diameter \mbox{$D_o = 206~\mathrm{\mu m}$}, relative velocity \mbox{$U_\mathrm{rel}= 6.45~\mathrm{m~s^{-1}}$} impact parameter \mbox{$X = 0.039$}, viscosity \mbox{$\mu_i= 5.55~\mathrm{mPas}$}, viscosity \mbox{$\mu_o = 19.15~\mathrm{mPas}$}, density \mbox{$\rho_i=1122.7~\mathrm{kg~m^{-3}}$}, density \mbox{$\rho_o=949.0~\mathrm{kg~m^{-3}}$}, surface tension \mbox{$\sigma_i= 68.3~\mathrm{mN~m^{-1}}$}, surface tension \mbox{$\sigma_o= 20.5~\mathrm{mN~m^{-1}}$}, interfacial tension \mbox{$\sigma_{io}= 35~\mathrm{mN~m^{-1}}$}. Simulation only: viscosity of air \mbox{$\mu_\mathrm{air} = 0.01824~\mathrm{mPas}$}, density of air \mbox{$\rho_\mathrm{air} = 1.19~\mathrm{kg~m^{-3}}$}.}  
\label{Fig:ValidationPicturesSingleReflexSOM20}

\end{figure*}  
\begin{figure*}[!tb]
 \includegraphics[page=3,scale=1,trim=0cm 1.4cm 0cm 0cm,clip]{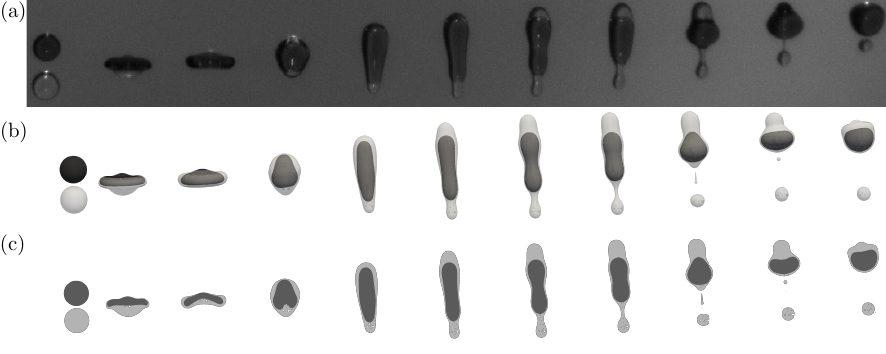} %
 \caption{Reflexive separation:  Comparison of (a) experiment and (b) outer view from simulations and (c) a section at the symmetry plane of the simulation. Collision of G50 inner (dark grey) and Perfluo outer liquid (transparent grey): droplet's diameter \mbox{$D_i = 199~\mathrm{\mu m}$}, droplet's diameter \mbox{$D_o = 197~\mathrm{\mu m}$}, relative velocity \mbox{$U_\mathrm{rel}= 3.17~\mathrm{m~s^{-1}}$}, impact parameter \mbox{$X = 0.018$}, viscosity \mbox{$\mu_i= 6.0~\mathrm{mPas}$}, viscosity \mbox{$\mu_o = 5.5~\mathrm{mPas}$}, density \mbox{$\rho_i=1126~\mathrm{kg~m^{-3}}$}, density \mbox{$\rho_o=1934.9~\mathrm{kg~m^{-3}}$}, surface tension \mbox{$\sigma_i= 68.6~\mathrm{mN~m^{-1}}$}, surface tension \mbox{$\sigma_o= 17.8~\mathrm{mN~m^{-1}}$}, interfacial tension \mbox{$\sigma_{io}= 36.5~\mathrm{mN~m^{-1}}$}. Simulation only: viscosity of air \mbox{$\mu_\mathrm{air} = 0.01824~\mathrm{mPas}$}, density of air \mbox{$\rho_\mathrm{air} = 1.19~\mathrm{kg~m^{-3}}$}.}
\label{Fig:ValidationPicturesReflexive}
\end{figure*}

\subsection{Experiments} \label{Sec:Experiments}
This section first presents the experimental set-up. Secondly, it describes the performed data analysis. 
Finally, the section ends with the used liquid properties and provides an overview of the experimentally probed parameter space. %
\subsubsection{Set-up}
Experiments were performed to produce head-on collisions of two quasi equally-sized droplets made of immiscible liquids. %
The set-up, which was described in detail elsewhere \cite{Planchette2009, Planchette2017}, consists of two droplet generators supplied each by one of the two immiscible liquids stored in independent pressurized tanks. %
The droplets production is achieved by forcing the Plateau-Rayleigh instability at a desired frequency, $f$, which is also used to drive a stroboscopic illumination. The stroboscopic illumination provides standing images made of several tens of superimposed collisions. %
External disturbances (air flow, vibrations,...) may cause small variations in the droplet trajectories, possibly giving a slightly blurry aspect to some pictures. %
This difficulty can be overcome by a second type of illumination producing single ultra short flashes (Nanolite). %
In both cases, a standard camera with a resolution of {4$~\mathrm{\mu m/\text{pixel}}$} records the collisions. Using this set-up, two streams of  droplets of approximately 200~$\mu m$ diameter  were produced and their respective trajectory adjusted to obtain (quasi) head-on collisions. To do so, the droplet generators were mounted onto translation and rotation micro-stages. The eccentricity of the collisions was controlled by optical observation in the plane of the collision and perpendicularly to it. The relative velocity of the droplets was adjusted to be close to the fragmentation threshold by varying the pressure in the tanks, the frequency of the droplet generators and the angle made by the two droplet streams in the collision plane. Typically, the relative velocity of the produced droplets ranged between 2 and 7~$\mathrm{m~s^{-1}}$.

\subsubsection{Type of data and their analysis}
The experimental approach provides images of the collisions showing the interaction of successive droplet pairs, cf.\ the illustrative pictures in Figs. \ref{Fig:ValidationPicturesCrossing}, \ref{Fig:ValidationPicturesSingleReflex} and \ref{Fig:ValidationPicturesReflexive}. 
These images are analysed using ImageJ \cite{ImageJ} providing  the diameter and center of mass of each droplet. %
Using $\vec{l}_{i}$ (and $\vec{l}_{o}$), the vector separating two successive drops of the same stream prior to collision and knowing $f$, the frequency at which the droplets are produced, we calculate their velocity as  $ \vec{U}_{i}=f \vec{l_{i}}$ (and $\vec{U}_{o}=f \vec{l_{o}}$) from which we deduce  $\vec{U}_\mathrm{rel}$ and $X$. %
{Throughout this work the relative velocity $U_\mathrm{rel} = |\vec{U}_\mathrm{rel}| = |\vec{U}_{i}-\vec{U}_{o}|$.} %
\citet{Planchette2009} provide further details. %
\\
It is important to note that the two immiscible liquids can be distinguished thanks to the usage of a dye, which is solely soluble in the inner liquid. %
As a result, the latter appears dark (grey) while the outer liquid remains transparent. %
Thus, a simple visual inspection of the recorded images indicates which one of the three outcomes - namely crossing separation, single reflex or reflexive separation, cf.\ Fig.~\ref{Fig:OutcomeSketch} occurs. %
Nevertheless, note that the precise spatial distribution of the two liquids cannot be obtained. %
Indeed only the projection of it into the collision plane is accessible.

\subsubsection{Used liquids and their properties}
The used immiscible liquids are of two types. The inner liquid is always made of a mixture of glycerol and water and referred to as GXX where XX indicates the weight percentage of glycerol. %
By varying the relative amount of both components, {the} viscosity can be adjusted from typically $1$ to $10~\mathrm{mPas}$ while keeping the density and surface tension constant at first order. %
The outer liquid is made either of silicon oil, SOMYY, whose viscosity (YY in mPa.s) ranges from $1.5$ to almost $20~\mathrm{mPas}$ for a density of approximately $900~\mathrm{kg~m^{-3}}$. %
Additionally, perfluorodecaline oil (Perfluo) offers a viscosity of $5~\mathrm{mPas}$ and a density of more than $1900~\mathrm{kg~m^{-3}}$. %
All oils provide full wetting when combined with the inner droplet's liquids. %
The spreading parameter varies only moderately ranging from $13$ to $16~\mathrm{mN~m^{-1}}$. %
All liquid parameters were measured in house using standard techniques: %
The density was obtained by weighing a known volume in a volumetric flask, the viscosity was measured using capillary viscosimeters and the surface and interfacial tensions were deduced from the analysis of pendant drops. %
The measured values are listed in the Table \ref{tab:SimExpSetup}, which summarizes the range covered by both the experimental and numerical method. %
Note that in addition to the physical quantities and liquid properties, the {droplets'} Weber and Reynolds numbers are given. %
Their respective definition reads: $We=\rho D {U_\mathrm{rel}}^2/\sigma$ and $Re=\rho D {U_\mathrm{rel}}/\mu$, where the inner/outer liquid properties are chosen for the inner/outer drop. %

\FloatBarrier
\subsection{Simulations} \label{Sec:Simulations}
Direct Numerical Simulations (DNS) were conducted
to explore parameter ranges and combinations, which are not fully accessible in the laboratory or to fill gaps between experimentally investigated cases. These are large variations of the surface and interfacial tensions and the spreading parameter as well as the droplet sizes. The simulation results thus enhance the basis for the analysis leading to the proposed model and its validation, cf.\ Sec.~\ref{Sec:Discussion}. Larger ranges of viscosities and viscosity ratios, on the other hand, were explored experimentally solely, because they pose a challenge for the simulations due to additional time-step constraints, which significantly increase the run-time. Thus, the simulations and experiments complement each other in order to broaden the range of explored parameters, cf. Tab.~\ref{tab:SimExpSetup}. %

In the following, first the simulation software is introduced briefly, then, the setup of the numerical simulations is described and finally, the kind of data evaluated from the simulations' results is discussed. %
Section~\ref{Sec:Validation} presents a validation with the previously described experiments in Sec.~\ref{Sec:Experiments}. \\

\subsubsection{Simulation software}\label{Sec:SimulationSoftware}
The DNS were performed with the multi-phase flow software Free Surface 3D (FS3D) \cite{Eisenschmidt2016}, which solves the incompressible Navier-Stokes equations in an one-field formulation on a Cartesian grid. %
FS3D employs a Volume of Fluid (VOF) method \cite{Hirt1981} with a Piecewise Linear Interface Calculation (PLIC)\cite{Youngs1982, Rider1998} for sharp interface reconstruction. %
As only one set of conservation equations needs to be solved for all phases, two indicator functions track the distribution of the disperse phases throughout the computational domain by means of the volume fractions.

The extension of this method to the interaction of two immiscible liquids in a gaseous environment was implemented and validated recently\cite{Potyka2023}. %
It includes an enhancement of the PLIC reconstruction and VOF advection by combining existing and new methods for the geometry reconstruction and transport in three-phase cells \cite{Potyka2023,Kromer2023}. %
Furthermore, an enhanced three-phase Continous Surface Stress (CSS) model \cite{Lafaurie1994} accounts for the three different surface and interfacial tensions acting at the three different interfaces. 
\\
The software FS3D is highly parallelized using Message Passing Interface (MPI) for a domain decomposition. This allowed running the DNS on the supercomputer HPE Apollo (Hawk) at the High Performance Computing Center at the University of Stuttgart (HLRS) on up to 1024 CPU cores. %

\subsubsection{Simulation setup}\label{SimulationSetup}
\begin{figure}[!htbp]
\def\svgwidth{8.5cm}
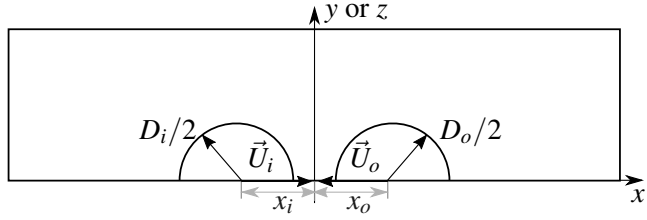
\caption{Sketch of the simulation setup: %
Two symmetry conditions are used to simulate a quarter of the droplet collision. As head-on collisions are considered (b=0), the $x$-$y$ and $x$-$z$ plane are identical and therefore only one plane is shown. %
}\label{Fig:SetupSim}
\end{figure}

The droplets' centers were initialized at coordinates \mbox{$x_o = 1.1 D_o$} and \mbox{$x_i = -1.1 D_i$} with respect to the plane at $x=0$. %
The simulation setup that considers two symmetry planes for the investigated head-on collisions is depicted in Fig.~\ref{Fig:SetupSim}. %
The diameters of the droplets, the inner $D_i$ and outer $D_o$, are the same as measured in the experiments and set exactly equal to an arbitrary value for the additional cases, which enlarge the parameter range investigated. %
Therefore, a unity diameter ratio is present in all simulation cases except for the  cases aiming to reproduce the experimental data, where the values measured in the experiments were employed. Similarly, when $X \neq 0$, the simulation set-up is modified to account for the loss of one symmetry plane. In this study, the diameters are varied from one simulation case to another ranging from $100$ to $800~\mathrm{\mu m}$.
The velocities $\vec{U}_\mathrm{i}$ and $\vec{U}_\mathrm{o}$ were chosen such, that the droplet collision's complex stays in the center of the computational domain. %
Thus, their initialized velocity value may deviate from $U_\mathrm{rel}/2$ to compensate especially for the density ratio. %
This is required to avoid much larger domains which cannot be computed in reasonable time on the supercomputer. %
The resolution with an equidistant Cartesian grid was adapted to the requirements of each case. %
Similarly, the size of the domain was adjusted for each case, such that also a larger collision complex fits the computational domain. %
In practice, resolutions from $0.78125$ to $12.5~\mathrm{\mu m/\text{cell}}$, i.e., $48$ to $144~\mathrm{\text{cells}/D}$, were sufficient depending on each combination of properties, droplet diameters, and relative velocity at the separation threshold.
The simulations with these resolutions shown in this study ran on 1 to 8 nodes of the supercomputer HPE Appollo (Hawk), with 128 CPU cores each, thus with up to 1024 processors in parallel. %
Table~\ref{tab:SimExpSetup} shows the ranges of the parameters initialized for the different simulations. %
The dataset on the Data Repository of the University of Stuttgart \citep{DarusData2023}  provides the full list of simulation setups for all cases shown in this work. %

\subsubsection{Obtained data from the simulations}
The simulations provide time-resolved, three dimensional information on the volume fraction distribution throughout the collision. %
The three-dimensional HDF5 data was converted into three-dimensional visualizations in Fig.s~\ref{Fig:ValidationPicturesCrossing}(b), \ref{Fig:ValidationPicturesSingleReflex}(b), and \ref{Fig:ValidationPicturesReflexive}(b) by the TPF Paraview Plugin \cite{tpfPlugin} for three-phase PLIC visualization developed by the Visualization Institute at the University of Stuttgart (VISUS) for this purpose. %
However, the simulation results allow a view on the liquids' distributions not only from the outside, but also from the inside of the droplet collision complex, e.g., Fig.~\ref{Fig:ValidationPicturesCrossing}(c), 
Fig.~\ref{Fig:ValidationPicturesSingleReflex}(c), and 
Fig.~\ref{Fig:ValidationPicturesReflexive}(c) for a section at the symmetry plane. %
These sections were extracted from the HDF5 output of FS3D with the aid of h5py \cite{h5py} and image processing tools provided by skimage \cite{skimage} and scipy \cite{scipy} in a python program. %
The maximum disc diameter is evaluated during the simulation in the simulation tool FS3D itself. 
The combination of the outside and inside view provides a more comprehensive overview of the liquids' motion during the immiscible-liquids' droplet collision. %
This aids in finding the relevant quantities and thus in identifying the competing driving mechanisms whose relative importance fixes the  collision outcomes. This basic analysis performed on our consolidated data-set has enabled the formulation of our model, cf.\ \ref{Sec:Discussion}. %

\subsubsection{Comparison of numerical and experimental data} \label{Sec:Validation}
  The good agreement between visualized DNS results and experimentally obtained images is shown exemplary in Figs. \ref{Fig:ValidationPicturesCrossing}, \ref{Fig:ValidationPicturesSingleReflex}, \ref{Fig:ValidationPicturesSingleReflexSOM20} and \ref{Fig:ValidationPicturesReflexive} for all three different liquid distributions. 
  It is important to note that all the illustrative pictures displayed here  were obtained with single ultra short flashes. While they generally offer greater sharpness than the ones recorded with stroboscopic illumination, they do not allow for averaging slight deviations found between successive droplet pairs and especially the potential variations of the impact parameter around its zero value. This explains why - for a given picture - the compound droplet does no show a perfectly constant orientation. This effect, well visible on the third element of Fig. \ref{Fig:ValidationPicturesSingleReflexSOM20}, can be attributed to experimental noise and should  not be considered while evaluating the validity of simulation data, that are produced with a fixed value of $X$.

Figure~\ref{Fig:ValidationPicturesCrossing} illustrates crossing separation of droplets of SOM5 and G50. %
The morphology of simulation (b-c) and experiment (a) agrees very well. %
Single reflex separation is depicted in Fig.~\ref{Fig:ValidationPicturesSingleReflex} and \ref{Fig:ValidationPicturesSingleReflexSOM20} with G50 and SOM10 or SOM20, respectively. %
For the collisions with SOM10, the simulation  (b-c), agrees almost perfectly with the experiments (a). For SOM20, ignoring the slight variations of the compound orientation visible in  Fig.~\ref{Fig:ValidationPicturesSingleReflexSOM20} (a), a good agreement is found with the numerical results (b-c).
Finally, reflexive separation can be observed for collisions between Perfluo and G50 droplets, cf.\ Fig.~\ref{Fig:ValidationPicturesReflexive}. %
This comparison reveals small morphological differences, which may be partly attributed to the premature breakdown of the ligament in the simulations (cf.\ third compound from the right). Yet, a further increase in the resolution which can avoid this premature breakdown is infeasible within a reasonable amount of compute time. Furthermore, the slight discrepancy could also originate uncertainties on the measured liquid properties or any other experimental uncertainties. %
Overall, the size and shape of the resulting compound and pure droplet agree well with the experimental results for all cases. %

\citet{Potyka2023} present further validation of the simulation of the interaction of immiscible liquids in air with the simulation framework FS3D.

\FloatBarrier

\section{Discussion} \label{Sec:Discussion}

\subsection{Phenomenological observations}\label{subsec:phenomeno}
Binary head-on collisions can be described in two phases \cite{Planchette2017}. %
The first phase represents the extension of the disk-shaped entity up to the maximum diameter, $D_\mathrm{max}$, caused by inertial forces. %
Note that what is called here a disk for simplicity consists indeed of a thin lamella surrounded by a toroidal rim. %
In analogy to a spring system, this first phase can be referred to as compression phase. %
The second phase, the relaxation phase, is determined by restoring forces due to surface and interfacial tensions. %
The deformed entity relaxes into a transversely elongated shape, called cylinder here for concision, which finally ruptures in the case of separation.
 
The fragmentation is caused by an excessive elongation during the relaxation phase. %
Its onset is described similarly to the well known Plateau-Rayleigh criterion employed to characterize the stability of periodically disturbed infinite liquid cylinders \cite{Planchette2012}. %
Despite its extreme simplicity, the criterion proved to be robust and was successfully employed for collisions between two and three droplets using one or two liquids \cite{Planchette2017}. %
Thus, for binary collisions of immiscible liquid droplets, we can assume that the composition of the daughter drops emerging from head-on fragmentation is directly related to the liquid distribution in this compound cylinder. %
Furthermore, \citet{planchette_2012} showed that the kinetics of the relaxation phase result from the geometry of the complex at maximal radial extension and the relevant liquid properties. Therefore, it seems legitimate to orientate our investigation around the following two main questions. %
First, what is the  liquid distribution in the complex at maximal extension and what fixes it? Can potential differences be correlated to different liquid compositions of the daughter drops?  %
Second, how is the liquid distribution obtained at maximal extension modified during the relaxation toward an excessively long cylinder? Can this evolution be used to predict the type of head-on separation taking place?
%
\newcommand{\thisscalea}{0.18}
\newcommand{\thisscaleb}{0.1694}
\newcommand{\thisscalec}{0.216}
\begin{figure*}[!htbp]
\begin{minipage}{\linewidth}
  \newcommand{\thispath}{./figures/CasesInsideRegimes/ExampleCrossing/G50-SOM5sigma15-512_X0_vrel300cms_20220706/}
  \newcommand{\rotangle}{270}
  \flushleft
  (a)\\
  \centering
  \null\hfill
  \includegraphics[scale=\thisscalea, trim=3cm 0.5cm 1cm 0cm, angle=\rotangle, clip]{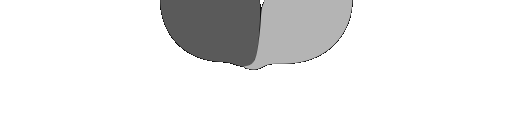}
  \hspace{-\removeindent}
  \reflectbox{\includegraphics[scale=\thisscalea, trim=3cm 0.5cm 1cm 0cm, angle=\rotangle, clip]{\thispath XYPlane_transparent-3.png}}
  \hfill
  \includegraphics[scale=\thisscalea, trim=3cm 0.5cm 1cm 0cm, angle=\rotangle, clip]{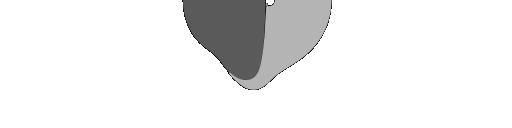}
  \hspace{-\removeindent}
  \reflectbox{\includegraphics[scale=\thisscalea, trim=3cm 0.5cm 1cm 0cm, angle=\rotangle, clip]{\thispath XYPlane_transparent-5.png}}
  \hfill
  \includegraphics[scale=\thisscalea, trim=3cm 0.5cm 1cm 0cm, angle=\rotangle, clip]{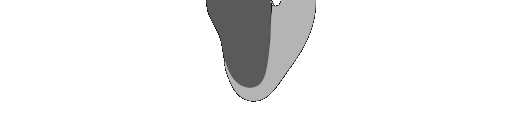}
   \hspace{-\removeindent}
   \reflectbox{\includegraphics[scale=\thisscalea, trim=3cm 0.5cm 1cm 0cm, angle=\rotangle, clip]{\thispath XYPlane_transparent-7.png}}
  \hfill
  \includegraphics[scale=\thisscalea, trim=3cm 0.5cm 1cm 0cm, angle=\rotangle, clip]{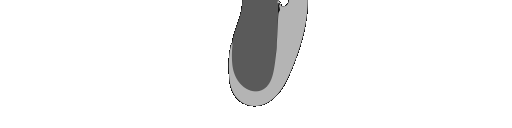}
   \hspace{-\removeindent}
      \reflectbox{ \includegraphics[scale=\thisscalea, trim=3cm 0.5cm 1cm 0cm, angle=\rotangle, clip]{\thispath XYPlane_transparent-9.png}}
  \hfill
  \includegraphics[scale=\thisscalea, trim=3cm 0.5cm 1cm 0cm, angle=\rotangle, clip]{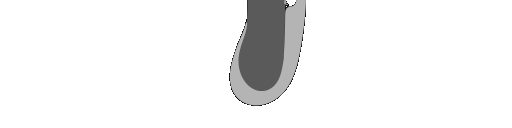}
   \hspace{-\removeindent}
    \reflectbox{ \includegraphics[scale=\thisscalea, trim=3cm 0.5cm 1cm 0cm, angle=\rotangle, clip]{\thispath XYPlane_transparent-11.png}}
    \hfill
  \includegraphics[scale=\thisscalea, trim=3cm 0.5cm 1cm 0cm, angle=\rotangle, clip]{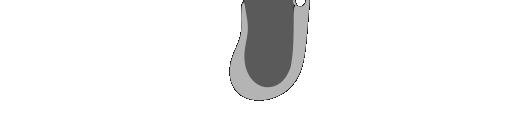}
   \hspace{-\removeindent}
    \reflectbox{ \includegraphics[scale=\thisscalea, trim=3cm 0.5cm 1cm 0cm, angle=\rotangle, clip]{\thispath XYPlane_transparent-13.png}}
  \hfill
   \includegraphics[scale=\thisscalea, trim=3cm 0.5cm 1cm 0cm, angle=\rotangle, clip]{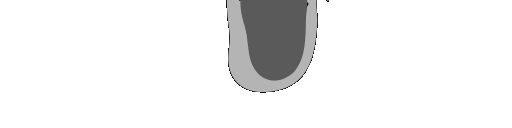}
   \hspace{-\removeindent}
    \reflectbox{ \includegraphics[scale=\thisscalea, trim=3cm 0.5cm 1cm 0cm, angle=\rotangle, clip]{\thispath XYPlane_transparent-15.png}}
  \hfill
  \rotatebox{180}{  
  \setlength{\unitlength}{1cm}
  \begin{picture}(0.1,2)
  \put(0,1.3){...}
  \end{picture}
  }
  \hfill
  \includegraphics[scale=\thisscalea, trim=3cm 0.5cm 1cm 0cm, angle=\rotangle, clip]{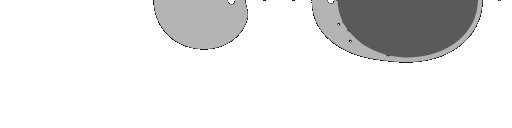}
  \hspace{-\removeindent}
  \reflectbox{\includegraphics[scale=\thisscalea, trim=3cm 0.5cm 1cm 0cm, angle=\rotangle, clip]{\thispath XYPlane_transparent-68.png}}
  \hfill\null
\end{minipage}
\begin{minipage}{\linewidth}
 \newcommand{\thispath}{./figures/CasesInsideRegimes/ExampleSingleReflex/G50-SOM20_768_X004_vrel645cms_20220706/}
  \newcommand{\rotangle}{270}
  \flushleft
  (b)\\
  \centering
  \null\hfill
  \includegraphics[scale=\thisscaleb, trim=7cm 0.5cm 9.5cm 0.5cm, angle=\rotangle, clip]{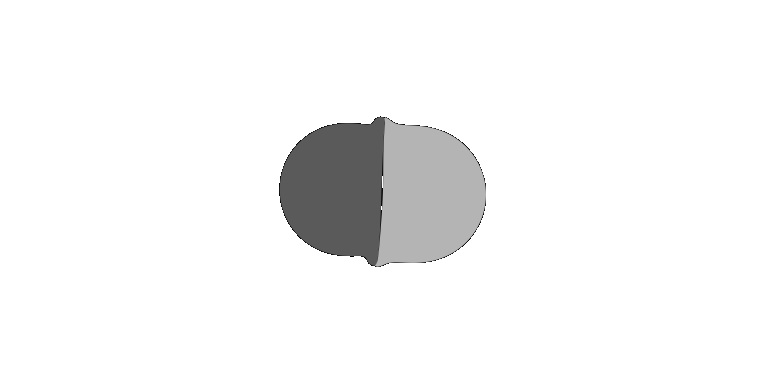}
  \hfill
  \includegraphics[scale=\thisscaleb, trim=7cm 0.5cm 9.5cm 0.5cm, angle=\rotangle, clip]{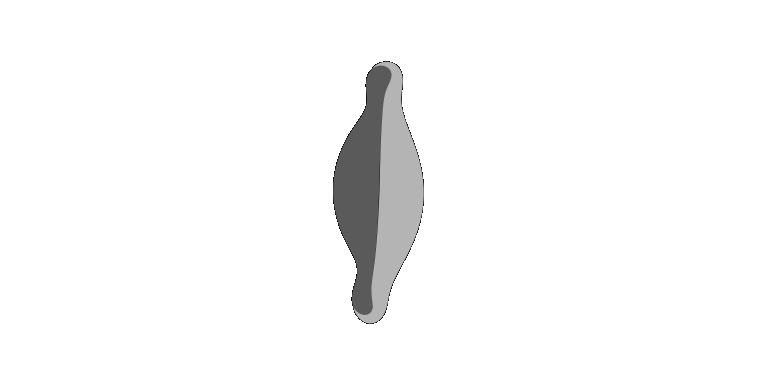}
  \hfill
  \includegraphics[scale=\thisscaleb, trim=7cm 0.5cm 9.5cm 0.5cm, angle=\rotangle, clip]{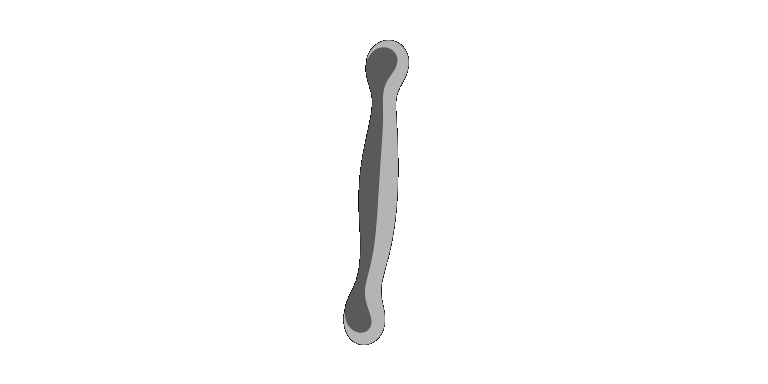}
  \hfill
  \includegraphics[scale=\thisscaleb, trim=7cm 0.5cm 9.5cm 0.5cm, angle=\rotangle, clip]{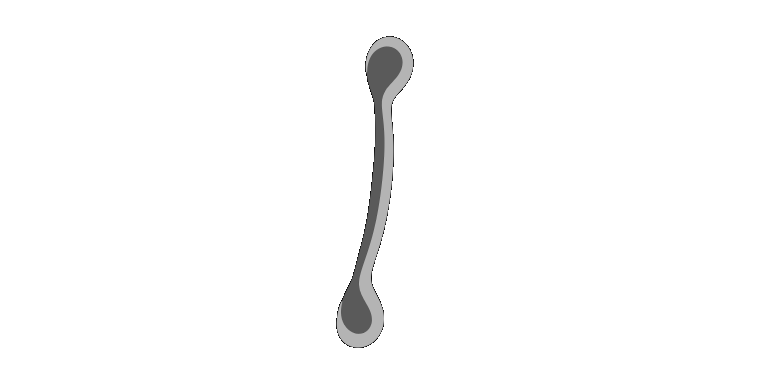}
  \hfill
  \includegraphics[scale=\thisscaleb, trim=7cm 0.5cm 9.5cm 0.5cm, angle=\rotangle, clip]{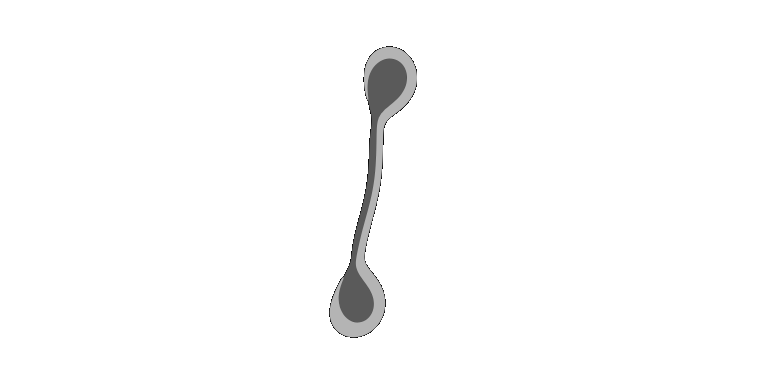}
  \hfill
  \includegraphics[scale=\thisscaleb, trim=7cm 0.5cm 9.5cm 0.5cm, angle=\rotangle, clip]{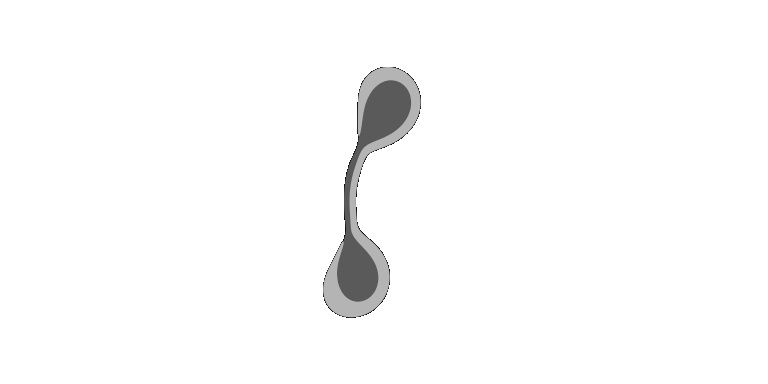}
  \hfill
  \rotatebox{180}{  
  \setlength{\unitlength}{1cm}
  \begin{picture}(0.1,2.4)
  \put(0,1.2){...}
  \end{picture}
  }
  \hfill
  \includegraphics[scale=\thisscaleb, trim=7cm 0.5cm 10cm 0.5cm, angle=\rotangle, clip]{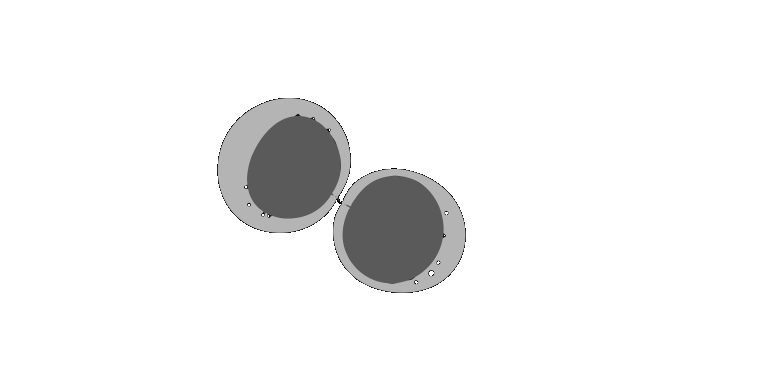}
  \hfill\null
\end{minipage}
\begin{minipage}{\linewidth}
  \newcommand{\thispath}{./figures/CasesInsideRegimes/ExampleReflexive/G50rho500-SOM5rho1000_768_X0_vrel500cms_20220706/}
  \newcommand{\rotangle}{270}
  \flushleft
  (c) \\
  \centering
  \null\hfill
  \includegraphics[scale=\thisscalec, trim=4cm 0cm 0cm 0cm, angle=\rotangle, clip]{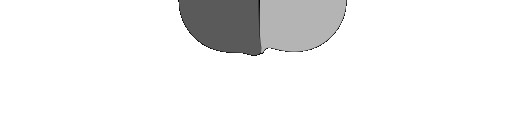}
  \hspace{-\removeindent}
  \reflectbox{\includegraphics[scale=\thisscalec, trim=4cm 0cm 0cm 0cm, angle=\rotangle, clip]{\thispath XYPlane_transparent-2.png}}
  \hfill
  \includegraphics[scale=\thisscalec, trim=4cm 0cm 0cm 0cm, angle=\rotangle, clip]{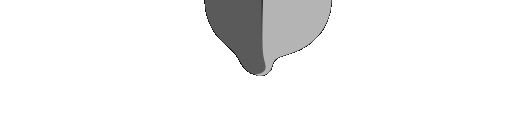}
  \hspace{-\removeindent}
  \reflectbox{\includegraphics[scale=\thisscalec, trim=4cm 0cm 0cm 0cm, angle=\rotangle, clip]{\thispath XYPlane_transparent-3.png}}
  \hfill
  \includegraphics[scale=\thisscalec, trim=4cm 0cm 0cm 0cm, angle=\rotangle, clip]{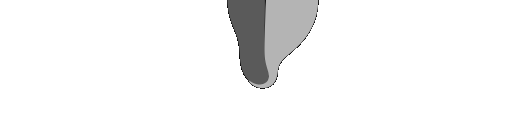}
   \hspace{-\removeindent}
   \reflectbox{\includegraphics[scale=\thisscalec, trim=4cm 0cm 0cm 0cm, angle=\rotangle, clip]{\thispath XYPlane_transparent-4.png}}
  \hfill
  \includegraphics[scale=\thisscalec, trim=4cm 0cm 0cm 0cm, angle=\rotangle, clip]{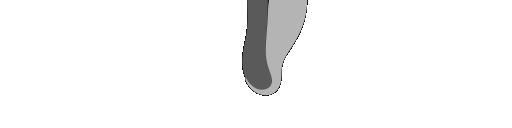}
   \hspace{-\removeindent}
      \reflectbox{ \includegraphics[scale=\thisscalec, trim=4cm 0cm 0cm 0cm, angle=\rotangle, clip]{\thispath XYPlane_transparent-5.png}}
  \hfill
  \includegraphics[scale=\thisscalec, trim=4cm 0cm 0cm 0cm, angle=\rotangle, clip]{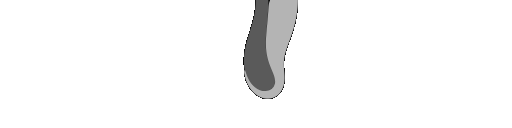}
   \hspace{-\removeindent}
    \reflectbox{ \includegraphics[scale=\thisscalec, trim=4cm 0cm 0cm 0cm, angle=\rotangle, clip]{\thispath XYPlane_transparent-6.png}}
    \hfill
  \includegraphics[scale=\thisscalec, trim=4cm 0cm 0cm 0cm, angle=\rotangle, clip]{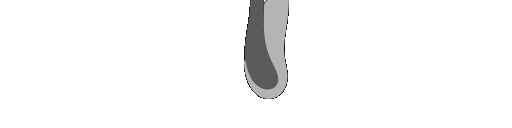}
   \hspace{-\removeindent}
    \reflectbox{ \includegraphics[scale=\thisscalec, trim=4cm 0cm 0cm 0cm, angle=\rotangle, clip]{\thispath XYPlane_transparent-7.png}}
     \hfill
  \includegraphics[scale=\thisscalec, trim=4cm 0cm 0cm 0cm, angle=\rotangle, clip]{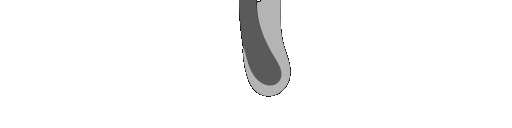}
   \hspace{-\removeindent}
    \reflectbox{ \includegraphics[scale=\thisscalec, trim=4cm 0cm 0cm 0cm, angle=\rotangle, clip]{\thispath XYPlane_transparent-8.png}}
  \hfill
  \rotatebox{180}{  
  \setlength{\unitlength}{1cm}
  \begin{picture}(0.1,2)
  \put(0,1.3){...}
  \end{picture}
  }
  \hfill
  \includegraphics[scale=\thisscalec, trim=4cm 0cm 0cm 0cm, angle=\rotangle, clip]{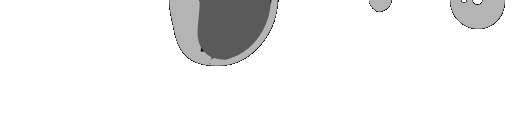}
  \hspace{-\removeindent}
  \reflectbox{\includegraphics[scale=\thisscalec, trim=4cm 0cm 0cm 0cm, angle=\rotangle, clip]{\thispath XYPlane_transparent-39.png}}
  \hfill\null
 \end{minipage}
\caption{Cross sections of the compression phase produced by numerical simulations with a time step $\Delta t$ between consecutive images. The last picture shows the collision outcome. (a) Crossing separation: $U_\mathrm{rel} = 3~\mathrm{m~s^{-1}}$, $D_o = D_i = 192~\mathrm{\mu m}$, $\rho_o = 913.4~\mathrm{kg~m^{-3}}$, $\rho_i = 1126.0 ~\mathrm{kg~m^{-3}}$, $\mu_o=4.57~\mathrm{mPas}$, $\mu_i=6.0 ~\mathrm{mPas}$, $\sigma_o =15.0 ~\mathrm{mN~m^{-1}}$, $\sigma_i = 68.6~\mathrm{mN~m^{-1}}$, $\sigma_{io} = 34.3~\mathrm{mN~m^{-1}}$, $\Delta t = 25~\mathrm{\mu s}$, (b) Single reflex: same parameters as in Fig.~\ref{Fig:ValidationPicturesSingleReflexSOM20} and $\Delta t = 35.5~\mathrm{\mu s}$, and (c) Reflexive separation: $U_\mathrm{rel} = 5~\mathrm{m~s^{-1}}$, $D_o = D_i = 200~\mathrm{\mu m}$, $\rho_o = 1000.0~\mathrm{kg~m^{-3}}$, $\rho_i = 500.0 ~\mathrm{kg~m^{-3}}$, $\mu_o=4.57~\mathrm{mPas}$, $\mu_i=6.0 ~\mathrm{mPas}$, $\sigma_o =19.5~\mathrm{mN~m^{-1}}$, $\sigma_i = 68.6~\mathrm{mN~m^{-1}}$, $\sigma_{io} = 34.3~\mathrm{mN~m^{-1}}$, $\Delta t = 17.5~\mathrm{\mu s}$. }\label{Fig:SimExampleCompression}   
\end{figure*}

\begin{figure*}[!htbp]
\begin{minipage}{\linewidth}
  \newcommand{\thispath}{./figures/CasesInsideRegimes/ExampleCrossing/G50-SOM5sigma15-512_X0_vrel300cms_20220706/}
  \newcommand{\rotangle}{270}
  \flushleft
  (a)\\
  \centering
  \null
  \includegraphics[scale=\thisscalea, trim=3cm 0.5cm 1cm 0cm, angle=\rotangle, clip]{\thispath XYPlane_transparent-3.png}
  \hspace{-\removeindent}
  \reflectbox{\includegraphics[scale=\thisscalea, trim=3cm 0.5cm 1cm 0cm, angle=\rotangle, clip]{\thispath XYPlane_transparent-3.png}}
  \hfill
 \rotatebox{180}{  
  \setlength{\unitlength}{1cm}
  \begin{picture}(0.1,2)
  \put(0,1.2){...}
  \end{picture}
  }
 \hfill
  \includegraphics[scale=\thisscalea, trim=3cm 0.5cm 1cm 0cm, angle=\rotangle, clip]{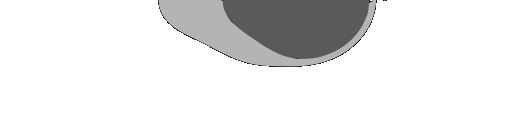}
   \hspace{-\removeindent}
    \reflectbox{\includegraphics[scale=\thisscalea, trim=3cm 0.5cm 1cm 0cm, angle=\rotangle, clip]{\thispath XYPlane_transparent-21.png}}
  \hfill
  \includegraphics[scale=\thisscalea, trim=3cm 0.5cm 1cm 0cm, angle=\rotangle, clip]{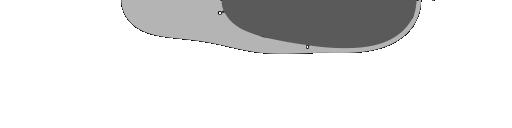}
  \hspace{-\removeindent}
  \reflectbox{\includegraphics[scale=\thisscalea, trim=3cm 0.5cm 1cm 0cm, angle=\rotangle, clip]{\thispath XYPlane_transparent-27.png}}
  \hfill
  \includegraphics[scale=\thisscalea, trim=3cm 0.5cm 1cm 0cm, angle=\rotangle, clip]{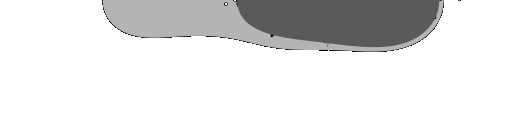}
   \hspace{-\removeindent}
   \reflectbox{\includegraphics[scale=\thisscalea, trim=3cm 0.5cm 1cm 0cm, angle=\rotangle, clip]{\thispath XYPlane_transparent-33.png}}
  \hfill
  \includegraphics[scale=\thisscalea, trim=3cm 0.5cm 1cm 0cm, angle=\rotangle, clip]{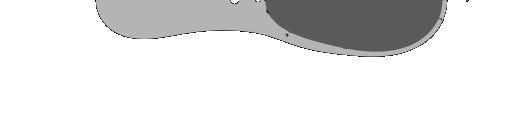}
   \hspace{-\removeindent}
      \reflectbox{ \includegraphics[scale=\thisscalea, trim=3cm 0.5cm 1cm 0cm, angle=\rotangle, clip]{\thispath XYPlane_transparent-39.png}}
  \hfill
  \includegraphics[scale=\thisscalea, trim=3cm 0.5cm 1cm 0cm, angle=\rotangle, clip]{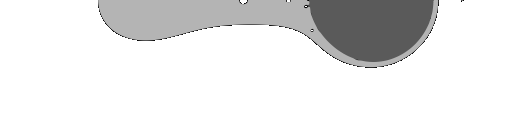}
   \hspace{-\removeindent}
    \reflectbox{ \includegraphics[scale=\thisscalea, trim=3cm 0.5cm 1cm 0cm, angle=\rotangle, clip]{\thispath XYPlane_transparent-45.png}}
  \hfill
  \includegraphics[scale=\thisscalea, trim=3cm 0.5cm 1cm 0cm, angle=\rotangle, clip]{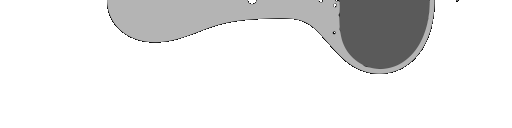}
   \hspace{-\removeindent}
    \reflectbox{\includegraphics[scale=\thisscalea, trim=3cm 0.5cm 1cm 0cm, angle=\rotangle, clip]{\thispath XYPlane_transparent-51.png}}
  \hfill
  \rotatebox{180}{  
  \setlength{\unitlength}{1cm}
  \begin{picture}(0.1,2)
  \put(0,1.2){...}
  \end{picture}
  }
  \hfill
  \includegraphics[scale=\thisscalea, trim=3cm 0.5cm 1cm 0cm, angle=\rotangle, clip]{\thispath XYPlane_transparent-68.png}
  \hspace{-\removeindent}
  \reflectbox{\includegraphics[scale=\thisscalea, trim=3cm 0.5cm 1cm 0cm, angle=\rotangle, clip]{\thispath XYPlane_transparent-68.png}}
  \null
\end{minipage}
\begin{minipage}{\linewidth}
  \newcommand{\thispath}{./figures/CasesInsideRegimes/ExampleSingleReflex/G50-SOM20_768_X004_vrel645cms_20220706/}
  \newcommand{\rotangle}{270}
  \flushleft
  (b)\\
  \centering
  \null
  \includegraphics[scale=\thisscaleb, trim=4cm 2.5cm 4cm 2.5cm, angle=\rotangle, clip]{\thispath XYPlane_transparent-3.png}
  \hfill
  \rotatebox{180}{  
  \setlength{\unitlength}{1cm}
  \begin{picture}(0.1,2)
  \put(0,1.6){...}
  \end{picture}
  }
  \hfill
  \includegraphics[scale=\thisscaleb, trim=4cm 2.5cm 4cm 2.5cm, angle=\rotangle, clip]{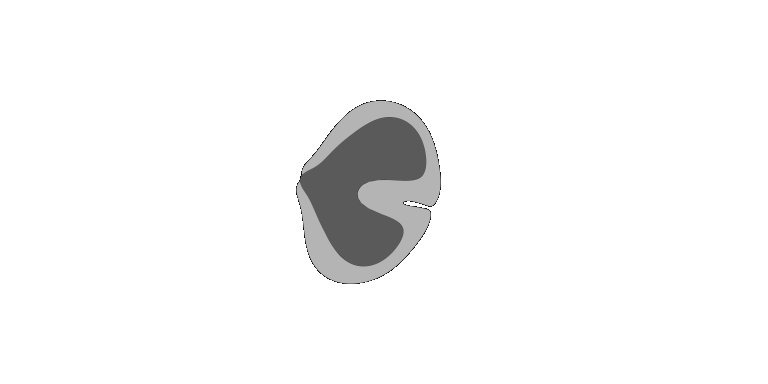}
  \hfill
  \includegraphics[scale=\thisscaleb, trim=4cm 2.5cm 4cm 2.5cm, angle=\rotangle, clip]{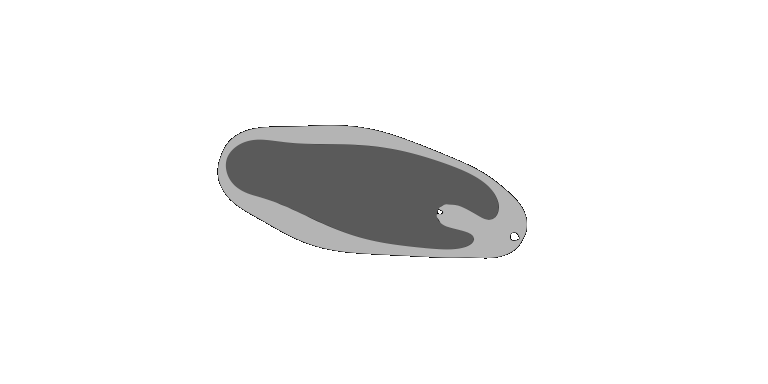}
  \hfill
  \includegraphics[scale=\thisscaleb, trim=4cm 2.5cm 4cm 2.5cm, angle=\rotangle, clip]{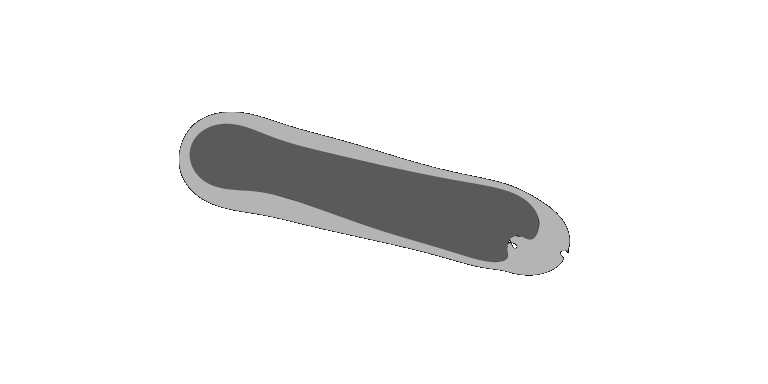}
  \hfill
   \includegraphics[scale=\thisscaleb, trim=4cm 2.5cm 4cm 2.5cm, angle=\rotangle, clip]{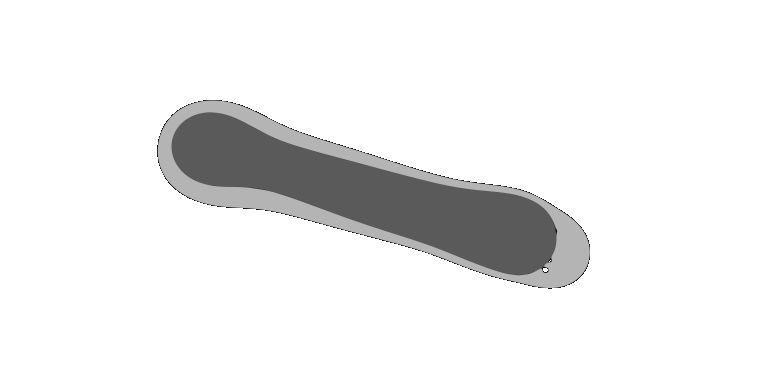}
  \hfill
   \includegraphics[scale=\thisscaleb, trim=4cm 2.5cm 4cm 2.5cm, angle=\rotangle, clip]{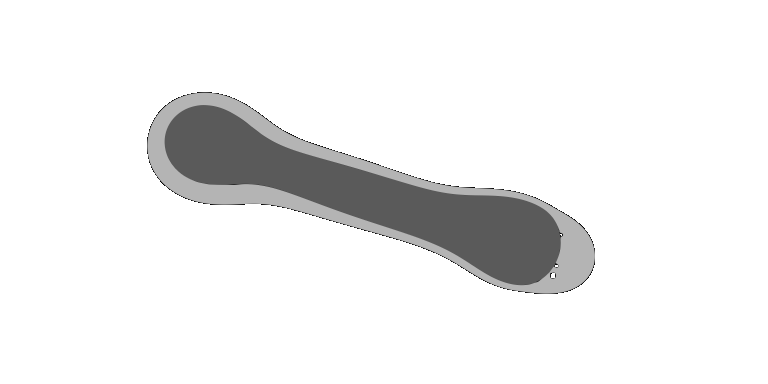}
  \hfill
  \includegraphics[scale=\thisscaleb, trim=4cm 2.5cm 4cm 2.5cm, angle=\rotangle, clip]{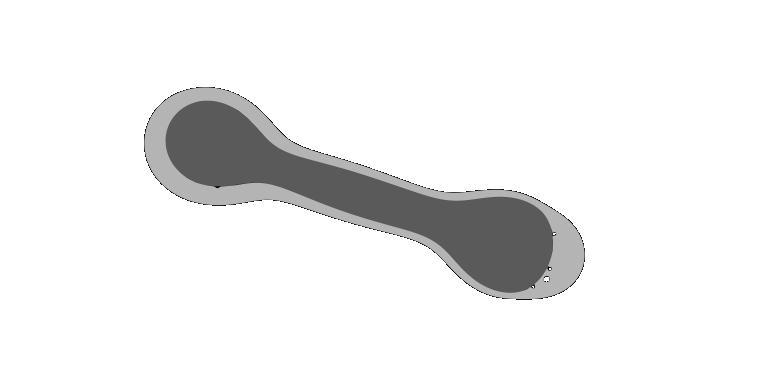}
  \hfill
   \includegraphics[scale=\thisscaleb, trim=4cm 2.5cm 4cm 2.5cm, angle=\rotangle, clip]{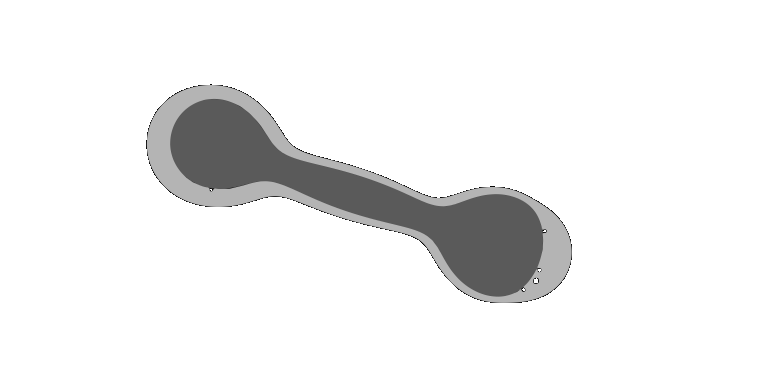}
  \hfill
  \rotatebox{180}{  
  \setlength{\unitlength}{1cm}
  \begin{picture}(0.1,2)
  \put(0,1.6){...}
  \end{picture}
  }
  \hfill
  \includegraphics[scale=\thisscaleb, trim=4cm 2.5cm 4cm 2.5cm, angle=\rotangle, clip]{\thispath XYPlane_transparent-100.png}
  \null
\end{minipage}
\begin{minipage}{\linewidth}
  \newcommand{\thispath}{./figures/CasesInsideRegimes/ExampleReflexive/G50rho500-SOM5rho1000_768_X0_vrel500cms_20220706/}
  \newcommand{\rotangle}{270}
  \flushleft
  (c)\\
  \centering
  \null
  \includegraphics[scale=\thisscalec, trim=4cm 0cm 0cm 0cm, angle=\rotangle, clip]{\thispath XYPlane_transparent-2.png}
  \hspace{-\removeindent}
  \reflectbox{\includegraphics[scale=\thisscalec, trim=4cm 0cm 0cm 0cm, angle=\rotangle, clip]{\thispath XYPlane_transparent-2.png}}
  \hfill
  \rotatebox{180}{  
  \setlength{\unitlength}{1cm}
  \begin{picture}(0.1,2)
  \put(0,1){...}
  \end{picture}
  }
  \hfill
  \includegraphics[scale=\thisscalec, trim=4cm 0cm 0cm 0cm, angle=\rotangle, clip]{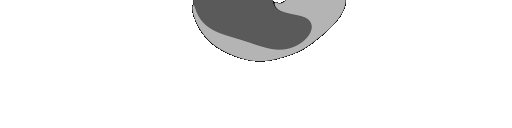}
  \hspace{-\removeindent}
  \reflectbox{\includegraphics[scale=\thisscalec, trim=4cm 0cm 0cm 0cm, angle=\rotangle, clip]{\thispath XYPlane_transparent-13.png}}
  \hfill
  \includegraphics[scale=\thisscalec, trim=4cm 0cm 0cm 0cm, angle=\rotangle, clip]{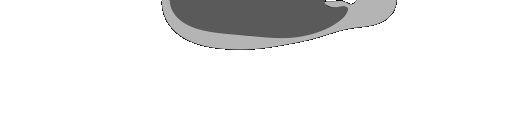}
   \hspace{-\removeindent}
   \reflectbox{\includegraphics[scale=\thisscalec, trim=4cm 0cm 0cm 0cm, angle=\rotangle, clip]{\thispath XYPlane_transparent-16.png}}
  \hfill
  \includegraphics[scale=\thisscalec, trim=4cm 0cm 0cm 0cm, angle=\rotangle, clip]{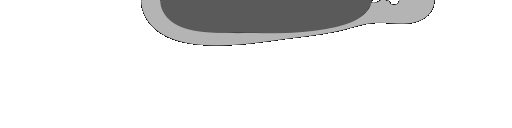}
   \hspace{-\removeindent}
      \reflectbox{ \includegraphics[scale=\thisscalec, trim=4cm 0cm 0cm 0cm, angle=\rotangle, clip]{\thispath XYPlane_transparent-19.png}}
  \hfill
  \includegraphics[scale=\thisscalec, trim=4cm 0cm 0cm 0cm, angle=\rotangle, clip]{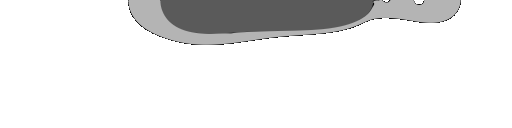}
   \hspace{-\removeindent}
    \reflectbox{ \includegraphics[scale=\thisscalec, trim=4cm 0cm 0cm 0cm, angle=\rotangle, clip]{\thispath XYPlane_transparent-22.png}}
    \hfill
  \includegraphics[scale=\thisscalec, trim=4cm 0cm 0cm 0cm, angle=\rotangle, clip]{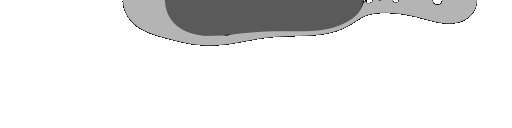}
   \hspace{-\removeindent}
    \reflectbox{ \includegraphics[scale=\thisscalec, trim=4cm 0cm 0cm 0cm, angle=\rotangle, clip]{\thispath XYPlane_transparent-25.png}}
     \hfill
  \includegraphics[scale=\thisscalec, trim=4cm 0cm 0cm 0cm, angle=\rotangle, clip]{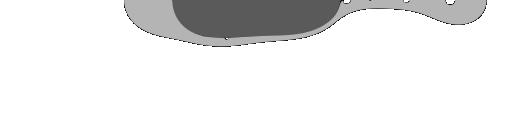}
   \hspace{-\removeindent}
    \reflectbox{ \includegraphics[scale=\thisscalec, trim=4cm 0cm 0cm 0cm, angle=\rotangle, clip]{\thispath XYPlane_transparent-28.png}}
  \hfill
  \rotatebox{180}{  
  \setlength{\unitlength}{1cm}
  \begin{picture}(0.1,2)
  \put(0,1){...}
  \end{picture}
  }
  \hfill
  \includegraphics[scale=\thisscalec, trim=4cm 0cm 0cm 0cm, angle=\rotangle, clip]{\thispath XYPlane_transparent-39.png}
  \hspace{-\removeindent}
  \reflectbox{\includegraphics[scale=\thisscalec, trim=4cm 0cm 0cm 0cm, angle=\rotangle, clip]{\thispath XYPlane_transparent-39.png}}
  \null
 \end{minipage}
 \caption{Cross sections of the relaxation phase produced by numerical simulations with a time step $\Delta t$ between consecutive images. The first (last) picture, shows the initial (final) liquid distribution. (a) Crossing separation: same parameters as in Fig.~\ref{Fig:SimExampleCompression}(a) with $\Delta t = 75~\mathrm{\mu s}$, (b) single reflex: same parameters as in Fig.~\ref{Fig:ValidationPicturesSingleReflexSOM20} and Fig.~\ref{Fig:SimExampleCompression}(b)  with $\Delta t = 71~\mathrm{\mu s}$, and (c)  reflexive separation: same parameters as in Fig.~\ref{Fig:SimExampleCompression}(c) with $\Delta t = 52.5~\mathrm{\mu s}$.}\label{Fig:SimExampleRelaxation}
\end{figure*}

A qualitative observation of the numerical simulations for the three regimes reveals  significant differences between the liquid distributions found close to maximal extension, i.e., at the end of the compression phase, cf.\ Fig.~\ref{Fig:SimExampleCompression}.

For crossing separation, by the time the collision complex reached its maximum diameter, $D_\mathrm{max}$, the outer liquid film covering the inner droplet is rather thick and its progression largely covers the side opposite to the impact. %
For the other two regimes, on the other hand, the majority of the outer fluid is still on the impact side. This seems to indicate that the liquid distribution at maximal disk extension fixes the possibility to obtain crossing separation (or not). During the relaxation phase, cf.\ Fig.~\ref{Fig:SimExampleRelaxation}(a), the outer liquid, which has mostly accumulated on the side opposite to the impact during compression phase gets elongated. %
The rupture of the collision complex results in a pure droplet of outer liquid that has migrated to the other side and a droplet of inner liquid that is encapsulated by a thin film. %
The inner droplet despite important deformation plays the role of a local obstacle. %
The momentum exchange between both droplets seems to be limited explaining the relative movement of the two liquids during the compression and relaxation phases. %
In the cases of reflexive and single reflex separation, however, the relaxation into a cylinder is initiated while a larger part of the outer liquid remains on the impact side, cf.\ Fig.~\ref{Fig:SimExampleRelaxation}(b) and Fig.~\ref{Fig:SimExampleRelaxation}(c). %
Further look into the relaxation phase evidences different behaviours for single reflex and reflexive separation, cf.\ Figs.~\ref{Fig:ValidationPicturesSingleReflex} -- \ref{Fig:ValidationPicturesReflexive} (c) and Fig.~\ref{Fig:SimExampleRelaxation}(b) and (c) respectively.
For single reflex separation, the outer envelop and the inner interface follow a comparable evolution. %
Both their shapes, and especially their lengths, but also their recoil kinetics are similar, leading to the formation of two droplets whose composition is at first order comparable. %
In contrast, for reflexive separation, the movement of the inner droplet seems relatively independent from the one of the outer liquid. %
The evolution of the two phases seem to decouple and while the inner liquid already recoiled into a sphere, the outer phase is still strongly elongated, forming protuberances at the compound extremities. %
Finally, the liquid protuberance at the impact side pinches-off and leaves a rather small droplet made purely of the outer phase, a topology called reflexive separation.

The liquid distribution developing during the compression phase differs for crossing separation from the ones observed for single reflex and reflexive separations. Hence, we first derive a criterion based on the compression phase to delimit this regime, cf.\ Sec.~\ref{subsec:discussion_crossing}. %
Single reflex and reflexive separation distinguish each other mostly during the relaxation of the compound entity, which shows different behaviours. %
It will therefore be treated in a second step, cf. Sec.~\ref{subsec:discussion_reflex}.
The two analyses are then put together to build a two-dimensional regime map, whose validity is tested using both experimental and numerical data, cf. Sec.~\ref{sec:visregimemap}.

\subsection{Distinguishing crossing separation from the other two regimes}\label{subsec:discussion_crossing}
%
\begin{figure*}[!tb]
    \centering
\begin{minipage}{\linewidth}
\flushleft
(a) \\
\centering
 \newcommand{\thispath}{./figures/RhoRatio/G50rho1000-SOM5rho500_768_X0_vrel450cms_20220706/}
  \newcommand{\thisheight}{1cm}
  \newcommand{\rotangle}{270}
  \null\hfill
  \includegraphics[height=\thisheight, trim=2cm 0cm 5cm 0cm, angle=\rotangle, clip]{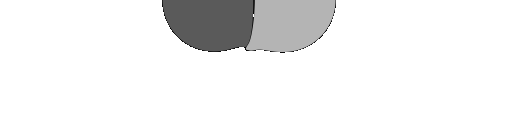}
  \hspace{-\removeindent}
  \reflectbox{\includegraphics[height=\thisheight, trim=2cm 0cm 5cm 0cm, angle=\rotangle, clip]{\thispath XYPlane_transparent-2.png}}
  \hfill
    \rotatebox{180}{  
  \setlength{\unitlength}{1cm}
  \begin{picture}(0.1,2)
  \put(0,1.5){...}
  \end{picture}
  }
  \hfill
  \includegraphics[height=\thisheight, trim=2cm 0cm 5cm 0cm, angle=\rotangle, clip]{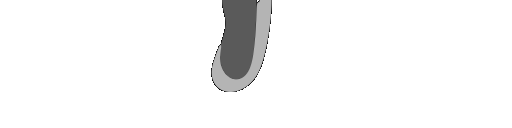}
   \hspace{-\removeindent}
    \reflectbox{ \includegraphics[height=\thisheight, trim=2cm 0cm 5cm 0cm, angle=\rotangle, clip]{\thispath XYPlane_transparent-8.png}}
  \hfill
   \includegraphics[height=\thisheight, trim=2cm 0cm 5cm 0cm, angle=\rotangle, clip]{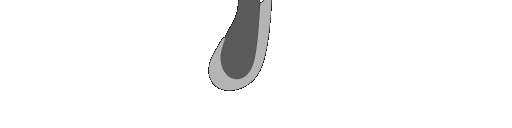}
   \hspace{-\removeindent}
    \reflectbox{ \includegraphics[height=\thisheight, trim=2cm 0cm 5cm 0cm, angle=\rotangle, clip]{\thispath XYPlane_transparent-9.png}}
  \hfill
   \includegraphics[height=\thisheight, trim=2cm 0cm 5cm 0cm, angle=\rotangle, clip]{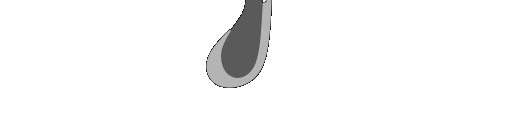}
   \hspace{-\removeindent}
    \reflectbox{ \includegraphics[height=\thisheight, trim=2cm 0cm 5cm 0cm, angle=\rotangle, clip]{\thispath XYPlane_transparent-10.png}}
  \hfill
     \includegraphics[height=\thisheight, trim=2cm 0cm 5cm 0cm, angle=\rotangle, clip]{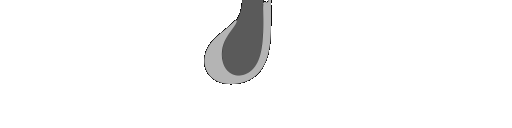}
   \hspace{-\removeindent}
    \reflectbox{ \includegraphics[height=\thisheight, trim=2cm 0cm 5cm 0cm, angle=\rotangle, clip]{\thispath XYPlane_transparent-11.png}}
  \hfill
     \includegraphics[height=\thisheight, trim=2cm 0cm 5cm 0cm, angle=\rotangle, clip]{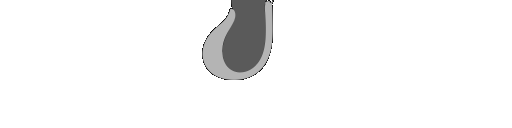}
   \hspace{-\removeindent}
    \reflectbox{ \includegraphics[height=\thisheight, trim=2cm 0cm 5cm 0cm, angle=\rotangle, clip]{\thispath XYPlane_transparent-12.png}}
  \hfill
  \rotatebox{180}{  
  \setlength{\unitlength}{1cm}
  \begin{picture}(0.1,2)
  \put(0,1.5){...}
  \end{picture}
  }
  \hfill
  \includegraphics[height=\thisheight, trim=2cm 0cm 5cm 0cm, angle=\rotangle, clip]{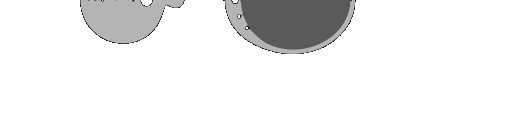}
  \hspace{-\removeindent}
  \reflectbox{\includegraphics[height=\thisheight, trim=2cm 0cm 5cm 0cm, angle=\rotangle, clip]{\thispath XYPlane_transparent-56.png}}
  \hfill\null
\end{minipage}   
\\
\begin{minipage}{\linewidth}
\flushleft
(b) \\
\centering
 \newcommand{\thispath}{./figures/RhoRatio/G50rho1000-SOM5rho1000_768_X0_vrel350cms_20220706/}
  \newcommand{\thisheight}{1cm}
  \newcommand{\rotangle}{270}
   \null\hfill
  \includegraphics[height=\thisheight, trim=5cm 0cm 5cm 0cm, angle=\rotangle, clip]{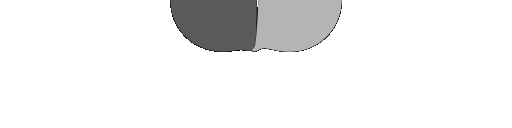}
  \hspace{-\removeindent}
  \reflectbox{\includegraphics[height=\thisheight, trim=5cm 0cm 5cm 0cm, angle=\rotangle, clip]{\thispath XYPlane_transparent-2.png}}
  \hfill
    \rotatebox{180}{  
  \setlength{\unitlength}{1cm}
  \begin{picture}(0.1,1.5)
  \put(0,1){...}
  \end{picture}
  }
  \hfill
  \includegraphics[height=\thisheight, trim=5cm 0cm 5cm 0cm, angle=\rotangle, clip]{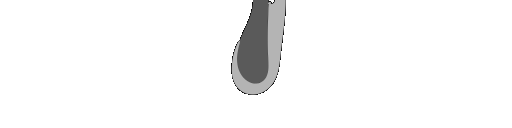}
   \hspace{-\removeindent}
    \reflectbox{ \includegraphics[height=\thisheight, trim=5cm 0cm 5cm 0cm, angle=\rotangle, clip]{\thispath XYPlane_transparent-8.png}}
  \hfill
   \includegraphics[height=\thisheight, trim=5cm 0cm 5cm 0cm, angle=\rotangle, clip]{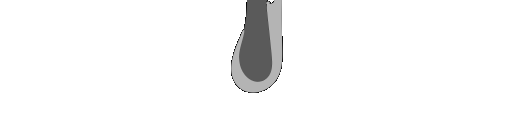}
   \hspace{-\removeindent}
    \reflectbox{ \includegraphics[height=\thisheight, trim=5cm 0cm 5cm 0cm, angle=\rotangle, clip]{\thispath XYPlane_transparent-9.png}}
  \hfill
   \includegraphics[height=\thisheight, trim=5cm 0cm 5cm 0cm, angle=\rotangle, clip]{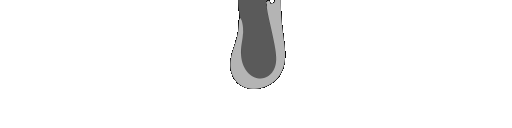}
   \hspace{-\removeindent}
    \reflectbox{ \includegraphics[height=\thisheight, trim=5cm 0cm 5cm 0cm, angle=\rotangle, clip]{\thispath XYPlane_transparent-10.png}}
  \hfill
     \includegraphics[height=\thisheight, trim=5cm 0cm 5cm 0cm, angle=\rotangle, clip]{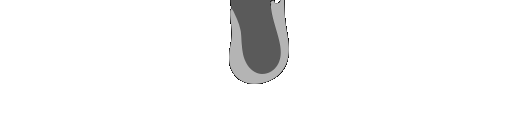}
   \hspace{-\removeindent}
    \reflectbox{ \includegraphics[height=\thisheight, trim=5cm 0cm 5cm 0cm, angle=\rotangle, clip]{\thispath XYPlane_transparent-11.png}}
  \hfill
     \includegraphics[height=\thisheight, trim=5cm 0cm 5cm 0cm, angle=\rotangle, clip]{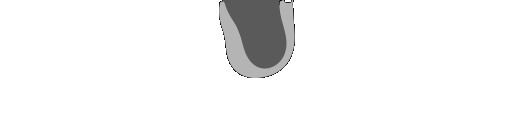}
   \hspace{-\removeindent}
    \reflectbox{ \includegraphics[height=\thisheight, trim=5cm 0cm 5cm 0cm, angle=\rotangle, clip]{\thispath XYPlane_transparent-12.png}}
  \hfill
  \rotatebox{180}{  
  \setlength{\unitlength}{1cm}
  \begin{picture}(0.1,1.5)
  \put(0,1){...}
  \end{picture}
  }
  \hfill
  \includegraphics[height=\thisheight, trim=5cm 0cm 5cm 0cm, angle=\rotangle, clip]{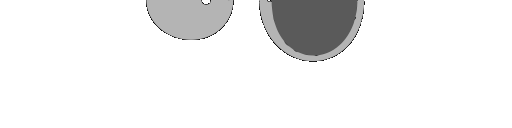}
  \hspace{-\removeindent}
  \reflectbox{\includegraphics[height=\thisheight, trim=5cm 0cm 5cm 0cm, angle=\rotangle, clip]{\thispath XYPlane_transparent-63.png}}
  \hfill\null
  \\
\end{minipage} 
\\
\begin{minipage}{\linewidth}
\flushleft
(c) \\
\centering
 \newcommand{\thispath}{./figures/RhoRatio/G50rho1000-SOM5rho2000_768_X0_vrel330cms_20220706/}
  \newcommand{\thisheight}{1cm}
  \newcommand{\rotangle}{270}
  \null\hfill
  \includegraphics[height=\thisheight, trim=1cm 0cm 0cm 0cm, angle=\rotangle, clip]{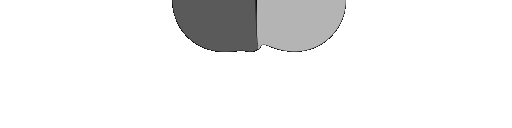}
  \hspace{-\removeindent}
  \reflectbox{\includegraphics[height=\thisheight, trim=1cm 0cm 0cm 0cm, angle=\rotangle, clip]{\thispath XYPlane_transparent-2.png}}
  \hfill
    \rotatebox{180}{  
  \setlength{\unitlength}{1cm}
  \begin{picture}(0.1,2)
  \put(0,1.7){...}
  \end{picture}
  }
  \hfill
  \includegraphics[height=\thisheight, trim=1cm 0cm 0cm 0cm, angle=\rotangle, clip]{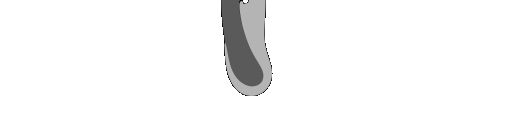}
   \hspace{-\removeindent}
    \reflectbox{ \includegraphics[height=\thisheight, trim=1cm 0cm 0cm 0cm, angle=\rotangle, clip]{\thispath XYPlane_transparent-8.png}}
  \hfill
   \includegraphics[height=\thisheight, trim=1cm 0cm 0cm 0cm, angle=\rotangle, clip]{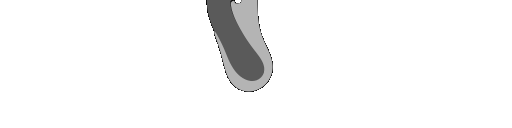}
   \hspace{-\removeindent}
    \reflectbox{ \includegraphics[height=\thisheight, trim=1cm 0cm 0cm 0cm, angle=\rotangle, clip]{\thispath XYPlane_transparent-9.png}}
  \hfill
   \includegraphics[height=\thisheight, trim=1cm 0cm 0cm 0cm, angle=\rotangle, clip]{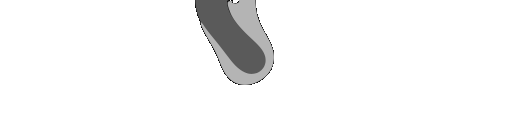}
   \hspace{-\removeindent}
    \reflectbox{ \includegraphics[height=\thisheight, trim=1cm 0cm 0cm 0cm, angle=\rotangle, clip]{\thispath XYPlane_transparent-10.png}}
  \hfill
     \includegraphics[height=\thisheight, trim=1cm 0cm 0cm 0cm, angle=\rotangle, clip]{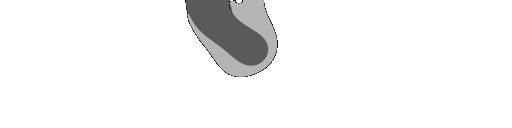}
   \hspace{-\removeindent}
    \reflectbox{ \includegraphics[height=\thisheight, trim=1cm 0cm 0cm 0cm, angle=\rotangle, clip]{\thispath XYPlane_transparent-11.png}}
  \hfill
     \includegraphics[height=\thisheight, trim=1cm 0cm 0cm 0cm, angle=\rotangle, clip]{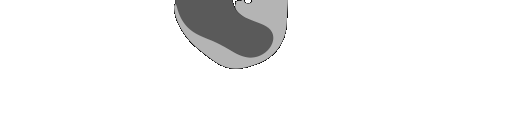}
   \hspace{-\removeindent}
    \reflectbox{ \includegraphics[height=\thisheight, trim=1cm 0cm 0cm 0cm, angle=\rotangle, clip]{\thispath XYPlane_transparent-12.png}}
  \hfill
    \rotatebox{180}{  
  \setlength{\unitlength}{1cm}
  \begin{picture}(0.1,2)
  \put(0,1.7){...}
  \end{picture}
  }
  \hfill
  \includegraphics[height=\thisheight, trim=1cm 0cm 0cm 0cm, angle=\rotangle, clip]{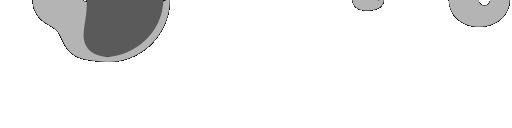}
  \hspace{-\removeindent}
  \reflectbox{\includegraphics[height=\thisheight, trim=1cm 0cm 0cm 0cm, angle=\rotangle, clip]{\thispath XYPlane_transparent-40.png}}
  \hfill\null
\end{minipage}
\caption{Visualisation of the lamella bending for different outer liquid's density and thus different relative velocities: (a) $\rho_o = 500~\mathrm{kg~m^{-3}}$ and $U_\mathrm{rel}=4.5~\mathrm{m~s^{-1}}$, (b) $\rho_o = 1000~\mathrm{kg~m^{-3}}$ and $U_\mathrm{rel}=3.5~\mathrm{m~s^{-1}}$ and (c) $\rho_o = 2000~\mathrm{kg~m^{-3}}$ and $U_\mathrm{rel}=3.3~\mathrm{m~s^{-1}}$. For all cases, the impact parameter is $X=0$,  the drop diameters and other liquid properties are kept constant: \mbox{$D_i = D_o = 200~\mathrm{\mu m}$},  $\rho_i=1000~\mathrm{kg~m^{-3}}$,  \mbox{$\mu_i= 4.57~\mathrm{mPas}$},   \mbox{$\sigma_i= 68.6~\mathrm{mN~m^{-1}}$}, \mbox{$\mu_o = 6.0~\mathrm{mPas}$}, \mbox{$\sigma_o= 19.5~\mathrm{mN~m^{-1}}$},  \mbox{$\sigma_{io}= 34.3~\mathrm{mN~m^{-1}}$.}}\label{Fig:RhoRatio}
\end{figure*}

Crossing separation can only occur if the outer liquid flowed sufficiently far around the inner droplet by the end of the compression phase. %
Thus, it is essential to estimate the progression of the outer liquid around the inner one and to relate it to the relevant dimension of the deformed compound. %
In a first step, we compare the progression of the outer liquid in the axial direction with the deformation of the compound droplet in radial direction. %
The outer liquid's progression is estimated by $L_{o}$, the distance covered by the macroscopic liquid film beyond the initial position of the  contact line in direction of impact. %
Assuming that the outer liquid flows around the inner one with a velocity scaling as their relative velocity, we obtain: $L_{o} \propto U_{rel} t_{inertia}$ where $t_{inertia}=D/U_{rel}$ is viewed as the time period during which momentum is exchanged along the axial direction. %
This analysis provides $L_{o} \propto D$.

We compare this to the radial extension caused by the deformation of the inner drop. %
Ignoring for now the flow of the outer liquid around the inner one, the contact line is expected to be found at a distance $D_\mathrm{max}/2$ of the compound center, where $D_\mathrm{max}$ is the maximal radial extension of the inner droplet. %
Several studies \cite{Planchette2012, Planchette2017, Baumgartner2020, Potyka2023} suggest that this extension remains independent from the outer liquid's properties and scales, at first  order, as $D_\mathrm{max} \propto D \sqrt{We_i}$. %
This result comes from the conversion of the not dissipated kinetic energy into surface energy during the compression phase, whereby the dissipated energy is a fixed fraction of the initial kinetic energy. %
Making use of this scaling, we obtain a first dimensionless parameter: 
\begin{equation}
\lambda = \frac{L_o}{D_\mathrm{max} }\sim  {We_i}^{-1/2}\text{.}
\end{equation}

At this stage of our analysis, the advection of the outer liquid by the deformation of the inner one, which is estimated by $D_\mathrm{max}$, is supposed to be purely radial. %
Inversely, the progression of the outer liquid on top of the inner one, which we called $L_o$, is considered to happen in the axial direction. %
In reality, in case the densities of the two liquids are different, the disk gets bent, which slightly modifies the ratio of axial and radial displacement, estimated so far by $\lambda$. %
This effect is visible on the sectional images from the simulations showing that for a varying  density ratio (cf.\ tab.~\ref{tab:SimExpSetup}), the inner droplet has different bending, cf. Fig.~\ref{Fig:RhoRatio}. %
We define this bending from the point of view of the outer liquid. %

A convex bending as in Fig.~\ref{Fig:RhoRatio}(a) is advantageous for the outer liquid's flow around the inner one. %
Indeed, the rim on which the outer liquid spreads is shifted toward the disk side where mostly the inner liquid is found. %
If, on the other hand, the bending is concave as in Fig.~\ref{Fig:RhoRatio}(c), the outer liquid's progression around the distorted inner drop is reduced by the slight backward motion of the rim, i.e., in the direction of the side where mostly outer liquid is found. %
The bending is a consequence of the momentum exchange between the two liquids along the relative velocity axis. %
For two droplets moving at $U_{rel}/2$, the denser droplet pushes the interface toward the less dense one, which explains why concave bending is found for $\rho_i < \rho_o$. %
As the process occurs continuously from the instant of contact and until the compound reaches its maximal extension, the central part of the disk is subjected to stronger shift than its periphery. %
The exact evaluation of the disk bending goes beyond the scope of our investigation. %
Here, we only aim at an appropriate correction of the proposed parameter $\lambda$, in the form of a new parameter $\Lambda=\lambda f(\theta)$ where $f$ is a simple function of $\zeta$, a relevant dimensionless parameter representing the axial and radial displacement due to the bending. %
Obvious dimensional analysis indicates that  $ \zeta= \rho_i / \rho_o$ is suitable. %
Further, the simplest function $f$ that  satisfies the required three conditions, namely: $f(1)=1$ indicating no bending for equal densities, $f(\zeta>1)>1$ and $f(\zeta<1)<1$ for the cases of convex and concave bendings respectively, is given by $f(\zeta)=\zeta=\rho_i / \rho_o$.

Putting these elements together, we finally obtain a dimensionless parameter, which is expected to distinguish  crossing separation from single reflex and reflexive separations. %
It reads: 
\begin{equation}
\Lambda = \frac{\rho_i}{\rho_o} \frac{1}{\sqrt{We_i}} 
\label{tau}
\end{equation}
We expect single reflex or reflexive separation at small and crossing separation at large values of $\Lambda$. This is tested against our experimental and numerical data in section \ref{sec:visregimemap}.

\subsection{Distinguishing single reflex from reflexive separation} \label{subsec:discussion_reflex}
Single reflex and reflexive separation differ qualitatively in whether the already encapsulated inner droplet neckles during the relaxation phase or the outer fluid breaks up after the encapsulation of the inner droplet. %
The distinction between the two fragmentation scenarios can be attributed to the relative importance of the viscous tangential stresses and the normal ones caused by surface and interfacial tensions that are acting. %

If the restoring forces on the encapsulated inner droplet are large compared to those acting on the outer envelop of the compound, the inner droplet recoils into a sphere almost independently from the simultaneous relaxation undergone at the level of the outer interface. %
Neglecting for now the tangential stresses, everything happens as if the inner droplet formed a 'hard' bead into a rather 'soft' liquid cylinder made of the outer liquid. %
Here, the adjective hard and soft refers to the Laplace pressure jump experienced for both phases. Such a configuration promotes the emergence of protuberances made of outer liquid, whose pinch-off leads to reflexive separation. %
Thus, a small ratio of the interfacial tension and the outer surface tension, $\sigma_{o}/\sigma_{io}$, is expected to cause reflexive separation. 

In order to better capture the relaxation phase, it is necessary to consider the tangential stresses as well. %
If the tangential stresses that act at the level of the inner droplet's surface are moderate, i.e., if the ratio of the kinematic viscosities $\frac{\nu_o}{\nu_i} = \frac{\mu_o / \rho_o}{\mu_i /\rho_i} $ is small, the motion of the inner liquid remains largely decoupled from the one of the outer liquid. %
As a consequence, reflexive separation occurs.

To complete our analysis, it may be interesting to focus on the opposite situation, i.e., when the inner and outer phases  recoil similarly, following the same shapes and kinetics, as typically observed for single reflex separation. %
For the two interfaces (between the inner liquid and the outer one, and between the outer liquid and the surrounding air) to develop similar curvatures,  the pressure jump associated to the outer interface must be significantly greater that the one related to the inner drop. %
Thus, this provides the condition  $\sigma_{o}/\sigma_{io}$ large for single reflex separation. %
In addition and as previously mentioned, the recoil of the inner drop  better follows the one of the outer liquid if the tangential stress this motion causes is large. %
This condition is given by  $\frac{\nu_o}{\nu_i}$ being large.

Combining the effects of the normal and tangential stresses, we build a dimensionless parameter, $N$, to distinguish the single reflex and reflexive separation. %
It reads:
\begin{equation}
N = \frac{\nu_o}{\nu_i}  \frac{\sigma_{o}}{\sigma_{io} }
\end{equation}
We expect single reflex to happen for large values of $N$ whereas the reflexive separation should be found for small $N$. %
The validity of our analysis is estimated in the next section where our criteria are tested against experimental and numerical results. %
%
\subsection{Visualization and test of the model}\label{sec:visregimemap}
%
\newcommand{\picwidth}{\linewidth}
\begin{figure*}
\newcommand{\plotthislambda}{lambda_mean}
\newcommand{\plotthislambdaerror}{lambda_error}
\newcommand{\plotfiglambda}{lambda}
\begin{tikzpicture}
\begin{axis}[
				axis lines=left,
				clip mode=individual,
				width=\picwidth,
				height=0.5*\picwidth,
				xmin=0,
			    xmax=0.72,
                ymin=0,
                ymax=6.2,
                xlabel={$\lambda$},
                ylabel={$N$},
                xtick distance = 0.1,                	
                ytick distance = 1,                	
		        line width=1.5pt,
		        legend style={
					at={(0.75,1.0)},
					anchor=north,
					legend columns=2,
					cells={anchor=west},
					rounded corners=4pt,}, 
				]

                \addplot[only marks,mark=square*,mark size=3 pt,color=red!60!black!60!, line width=1pt,error bars/.cd, x dir=both, x explicit] table[x=\plotthislambda,y=NU, x error=\plotthislambdaerror]{./data/Crossing_Regimemap_Experiments.txt};
                \addplot[only marks,mark=square,mark size=3 pt,color=red, line width=1pt,error bars/.cd, x dir=both, x explicit] table[x=\plotthislambda,y=NU, x error=\plotthislambdaerror] {./data/Crossing_Regimemap.txt};
                \addplot[only marks,mark=diamond*,mark size=4 pt,color=green!60!black!60!, line width=1pt,error bars/.cd, x dir=both, x explicit] table[x=\plotthislambda,y=NU, x error=\plotthislambdaerror]{./data/SingleReflex_Regimemap_Experiments.txt};
                \addplot[only marks,mark=diamond,mark size=4 pt,color=green, line width=1pt,error bars/.cd, x dir=both, x explicit] table[x=\plotthislambda,y=NU, x error=\plotthislambdaerror] {./data/SingleReflex_Regimemap.txt};
              \addplot[only marks,mark=triangle*,mark size=4pt,color=blue!60!black!60!, line width=1pt,error bars/.cd, x dir=both, x explicit] table[x=\plotthislambda,y=NU, x error=\plotthislambdaerror] {./data/Reflexive_Regimemap_Experiments.txt};               
              \addplot[only marks,mark=triangle,mark size=4 pt,color=blue, line width=1pt,error bars/.cd, x dir=both, x explicit] table[x=\plotthislambda,y=NU, x error=\plotthislambdaerror] {./data/Reflexive_Regimemap.txt};

             \addplot[only marks,mark=o,mark size=4.5 pt,color=black, line width=1pt] table[x=\plotthislambda,y=NU] {./data/PartialWetting_Regimemap.txt};
             \addplot[only marks,mark=x,mark size=3 pt,color=black, line width=1pt] table[x=\plotfiglambda,y=NU] {./data/Figures_Regimemap.txt};
             
             \node[font=\scriptsize] at (axis cs: 0.045,2.39) {\ref{Fig:ValidationPicturesSingleReflexSOM20}, \ref{Fig:SimExampleCompression} (b), \ref{Fig:SimExampleRelaxation} (b)};
             \node[font=\scriptsize] at (axis cs: 0.11,1.09) {\ref{Fig:ValidationPicturesSingleReflex}};
             
             \coordinate (insetPosition) at (rel axis cs:1,0.28);

        
\end{axis}
\begin{axis}[at={(insetPosition)},anchor={outer south east},footnotesize,
                clip mode=individual,
				width=9cm,
				height=7.2cm,
				xmin=0.091,
			    xmax=0.27,
                ymin=0,
                ymax=1.13,
                xtick distance = 0.03,                	
                ytick distance = 0.2,                	
		        line width=1.5pt, 
				]

                \addplot[only marks,mark=square*,mark size=3 pt,color=red!60!black!60!, line width=1pt,error bars/.cd, x dir=both, x explicit] table[x=\plotthislambda,y=NU, x error=\plotthislambdaerror]{./data/Crossing_Regimemap_Experiments.txt};
                \addplot[only marks,mark=square,mark size=3 pt,color=red, line width=1pt,error bars/.cd, x dir=both, x explicit] table[x=\plotthislambda,y=NU, x error=\plotthislambdaerror] {./data/Crossing_Regimemap.txt};
                \addplot[only marks,mark=diamond*,mark size=4 pt,color=green!60!black!60!, line width=1pt,error bars/.cd, x dir=both, x explicit] table[x=\plotthislambda,y=NU, x error=\plotthislambdaerror]{./data/SingleReflex_Regimemap_Experiments.txt};
                \addplot[only marks,mark=diamond,mark size=4 pt,color=green, line width=1pt,error bars/.cd, x dir=both, x explicit] table[x=\plotthislambda,y=NU, x error=\plotthislambdaerror] {./data/SingleReflex_Regimemap.txt};
              \addplot[only marks,mark=triangle*,mark size=4pt,color=blue!60!black!60!, line width=1pt,error bars/.cd, x dir=both, x explicit] table[x=\plotthislambda,y=NU, x error=\plotthislambdaerror] {./data/Reflexive_Regimemap_Experiments.txt};               
              \addplot[only marks,mark=triangle,mark size=4 pt,color=blue, line width=1pt,error bars/.cd, x dir=both, x explicit] table[x=\plotthislambda,y=NU, x error=\plotthislambdaerror] {./data/Reflexive_Regimemap.txt};

             \addplot[only marks,mark=o,mark size=4.5 pt,color=black, line width=1pt] table[x=\plotthislambda,y=NU] {./data/PartialWetting_Regimemap.txt};  
             \addplot[only marks,mark=x,mark size=3 pt,color=black, line width=1pt] table[x=\plotfiglambda,y=NU] {./data/Figures_Regimemap.txt};

             \node[font=\scriptsize] at (axis cs: 0.115,1.09) {\ref{Fig:ValidationPicturesSingleReflex}};
             \node[font=\scriptsize] at (axis cs: 0.146,0.627) {\ref{Fig:ValidationPicturesCrossing}};
             \node[font=\scriptsize] at (axis cs: 0.174,0.31) {\ref{Fig:ValidationPicturesReflexive}};
             \node[font=\scriptsize] at (axis cs: 0.193, 0.455) {\ref{Fig:SimExampleCompression}(a),\ref{Fig:SimExampleRelaxation}(a)};
             \node[font=\scriptsize] at (axis cs: 0.16, 0.16) {\ref{Fig:SimExampleCompression}(c),\ref{Fig:SimExampleRelaxation}(c)};
             \node[font=\scriptsize] at (axis cs: 0.12,0.866) {\ref{Fig:RhoRatio}(a)};
             \node[font=\scriptsize] at (axis cs: 0.16,0.397) {\ref{Fig:RhoRatio}(b)};
             \node[font=\scriptsize] at (axis cs: 0.18,0.16) {\ref{Fig:RhoRatio}(c)};
                          
    \end{axis}
\end{tikzpicture}
\caption{ ($\lambda$-$N$) regime map. Without accounting for the bending of the lamella via the liquid density ratio,  the crossing separation (red squares) cannot be well distinguished from the single reflex (green diamonds) and reflexive (blue triangles) separations. Empty (full) symbols represent simulation (experimental) data. Partial wetting is indicated by additional black circles.  The error bars indicate the highest velocity of merging and the lowest velocity of separation found for each fluid combination. The symbols in the center of each interval represent $U_\mathrm{rel}$ at the threshold of head-on separation. Inset: Zoom around the regime transitions (same data). The data points corresponding to figures of droplet collisions presented in this work are marked by black crosses with their respective figure number(s). 
Corresponding data can be found in a dataset on the Data Repository of the University of Stuttgart (DaRUS) \citep{DarusData2023}.}\label{Fig:RegimemaplambdaNoRho}
\end{figure*}
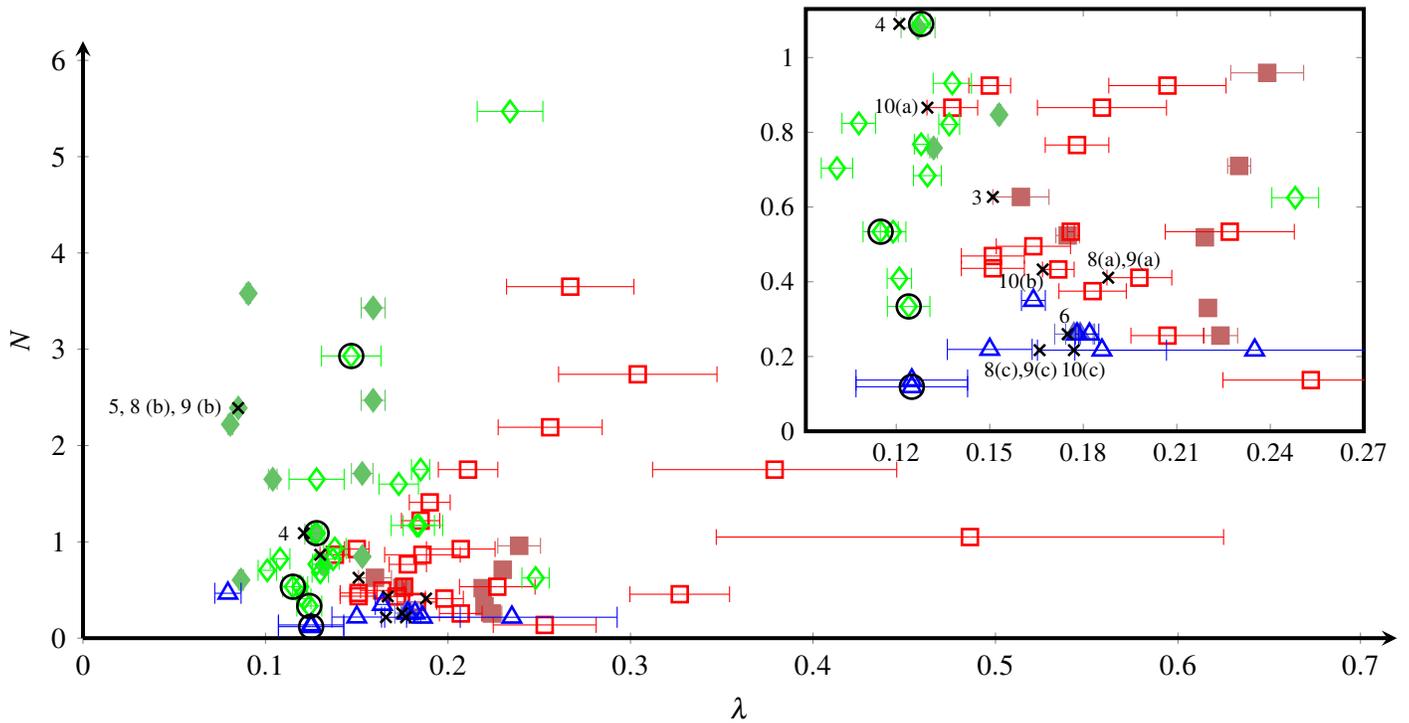
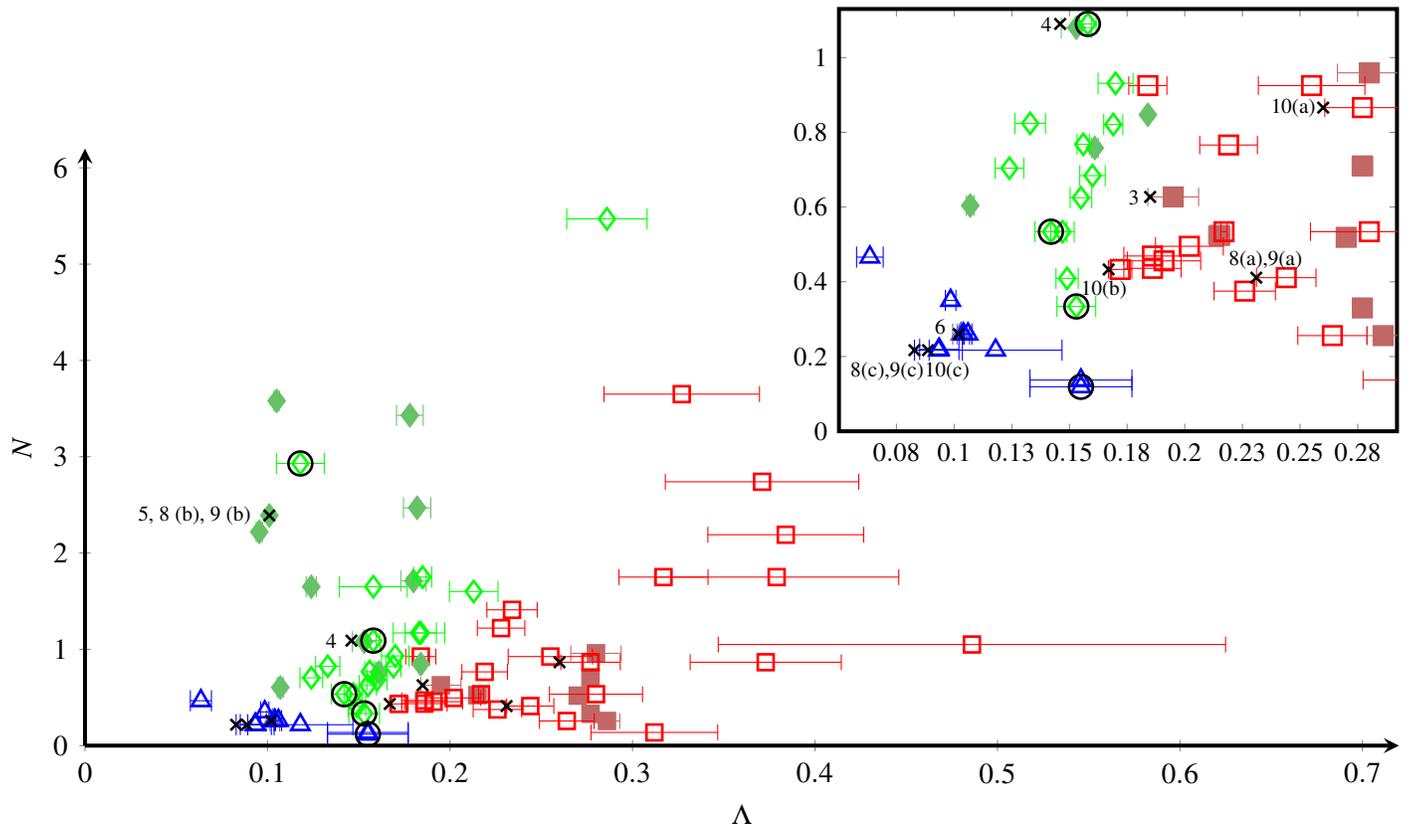
\begin{figure*}[!tb]
\newcommand{\plotthislambda}{Lambda_mean}
\newcommand{\plotthislambdaerror}{Lambda_error}
\newcommand{\plotfiglambda}{Lambda}

\begin{tikzpicture}
\begin{axis}[
				axis lines=left,
				clip mode=individual,
				width=\picwidth,
				height=0.5*\picwidth,
				xmin=0,
			    xmax=0.72,
                ymin=0,
                ymax=6.2,
                xlabel={$\Lambda$},
                ylabel={$N$},
                xtick distance = 0.1,                	
                ytick distance = 1,                	
		        line width=1.5pt,
		        legend style={
					at={(0.75,1.0)},
					anchor=north,
					legend columns=2,
					cells={anchor=west},
					rounded corners=4pt,}, 
				]

                \addplot[only marks,mark=square*,mark size=3 pt,color=red!60!black!60!, line width=1pt,error bars/.cd, x dir=both, x explicit] table[x=\plotthislambda,y=NU, x error=\plotthislambdaerror]{./data/Crossing_Regimemap_Experiments.txt};
                \addplot[only marks,mark=square,mark size=3 pt,color=red, line width=1pt,error bars/.cd, x dir=both, x explicit] table[x=\plotthislambda,y=NU, x error=\plotthislambdaerror] {./data/Crossing_Regimemap.txt};
                \addplot[only marks,mark=diamond*,mark size=4 pt,color=green!60!black!60!, line width=1pt,error bars/.cd, x dir=both, x explicit] table[x=\plotthislambda,y=NU, x error=\plotthislambdaerror]{./data/SingleReflex_Regimemap_Experiments.txt};
                \addplot[only marks,mark=diamond,mark size=4 pt,color=green, line width=1pt,error bars/.cd, x dir=both, x explicit] table[x=\plotthislambda,y=NU, x error=\plotthislambdaerror] {./data/SingleReflex_Regimemap.txt};
              \addplot[only marks,mark=triangle*,mark size=4pt,color=blue!60!black!60!, line width=1pt,error bars/.cd, x dir=both, x explicit] table[x=\plotthislambda,y=NU, x error=\plotthislambdaerror] {./data/Reflexive_Regimemap_Experiments.txt};               
              \addplot[only marks,mark=triangle,mark size=4 pt,color=blue, line width=1pt,error bars/.cd, x dir=both, x explicit] table[x=\plotthislambda,y=NU, x error=\plotthislambdaerror] {./data/Reflexive_Regimemap.txt};

             \addplot[only marks,mark=o,mark size=4.5 pt,color=black, line width=1pt] table[x=\plotthislambda,y=NU] {./data/PartialWetting_Regimemap.txt};
             \addplot[only marks,mark=x,mark size=3 pt,color=black, line width=1pt] table[x=\plotfiglambda,y=NU] {./data/Figures_Regimemap.txt};

             \node[font=\scriptsize] at (axis cs: 0.06,2.39) {\ref{Fig:ValidationPicturesSingleReflexSOM20}, \ref{Fig:SimExampleCompression} (b), \ref{Fig:SimExampleRelaxation} (b)};
             \node[font=\scriptsize] at (axis cs: 0.135,1.09) {\ref{Fig:ValidationPicturesSingleReflex}};

             
             \coordinate (insetPosition) at (rel axis cs:1,0.46);

        
\end{axis}
\begin{axis}[at={(insetPosition)},anchor={outer south east},footnotesize,
                clip mode=individual,
				width=9cm,
				height=7.2cm,
				xmin=0.0501,
			    xmax=0.292,
                ymin=0,
                ymax=1.13,
                xtick distance = 0.025,                	
                ytick distance = 0.2,     
                x tick label style={
                    /pgf/number format/.cd,
                    fixed,},
		        line width=1.5pt, 
				]

                \addplot[only marks,mark=square*,mark size=3.5 pt,color=red!60!black!60!, line width=1pt,error bars/.cd, x dir=both, x explicit] table[x=\plotthislambda,y=NU, x error=\plotthislambdaerror]{./data/Crossing_Regimemap_Experiments.txt};
                \addplot[only marks,mark=square,mark size=3.5 pt,color=red, line width=1pt,error bars/.cd, x dir=both, x explicit] table[x=\plotthislambda,y=NU, x error=\plotthislambdaerror] {./data/Crossing_Regimemap.txt};
                \addplot[only marks,mark=diamond*,mark size=4 pt,color=green!60!black!60!, line width=1pt,error bars/.cd, x dir=both, x explicit] table[x=\plotthislambda,y=NU, x error=\plotthislambdaerror]{./data/SingleReflex_Regimemap_Experiments.txt};
                \addplot[only marks,mark=diamond,mark size=4 pt,color=green, line width=1pt,error bars/.cd, x dir=both, x explicit] table[x=\plotthislambda,y=NU, x error=\plotthislambdaerror] {./data/SingleReflex_Regimemap.txt};
              \addplot[only marks,mark=triangle*,mark size=4pt,color=blue!60!black!60!, line width=1pt,error bars/.cd, x dir=both, x explicit] table[x=\plotthislambda,y=NU, x error=\plotthislambdaerror] {./data/Reflexive_Regimemap_Experiments.txt};               
              \addplot[only marks,mark=triangle,mark size=4 pt,color=blue, line width=1pt,error bars/.cd, x dir=both, x explicit] table[x=\plotthislambda,y=NU, x error=\plotthislambdaerror] {./data/Reflexive_Regimemap.txt};

             \addplot[only marks,mark=o,mark size=4.5 pt,color=black, line width=1pt] table[x=\plotthislambda,y=NU] {./data/PartialWetting_Regimemap.txt};  
             \addplot[only marks,mark=x,mark size=3 pt,color=black, line width=1pt] table[x=\plotfiglambda,y=NU] {./data/Figures_Regimemap.txt};

             \node[font=\scriptsize] at (axis cs: 0.14,1.09) {\ref{Fig:ValidationPicturesSingleReflex}};
             \node[font=\scriptsize] at (axis cs: 0.178,0.627) {\ref{Fig:ValidationPicturesCrossing}};
             \node[font=\scriptsize] at (axis cs: 0.094,0.28) {\ref{Fig:ValidationPicturesReflexive}};
             \node[font=\scriptsize] at (axis cs: 0.235, 0.46) {\ref{Fig:SimExampleCompression}(a),\ref{Fig:SimExampleRelaxation}(a)};
             \node[font=\scriptsize] at (axis cs: 0.071, 0.16) {\ref{Fig:SimExampleCompression}(c),\ref{Fig:SimExampleRelaxation}(c)};
             \node[font=\scriptsize] at (axis cs: 0.247,0.866) {\ref{Fig:RhoRatio}(a)};
             \node[font=\scriptsize] at (axis cs: 0.165,0.38) {\ref{Fig:RhoRatio}(b)};
             \node[font=\scriptsize] at (axis cs: 0.097,0.16) {\ref{Fig:RhoRatio}(c)};   
    \end{axis}
\end{tikzpicture}
\caption{ ($\Lambda$-$N$) regime map. With accounting for the bending of the lamella via the liquid density ratio, the crossing separation (red squares) can be well distinguished from the single reflex (green diamonds) and reflexive (blue triangles) separation results. Empty (full) symbols represent simulation (experimental) data. Partial wetting is indicated by additional black circles. The error bars indicate the highest velocity of merging and the lowest velocity of separation found for each fluid combination. The symbols in the center of each interval represent $U_\mathrm{rel}$ at the threshold of head-on separation. Inset: Zoom around the regime transitions (same data). The data points corresponding to figures of droplet collisions presented in this work are marked by black crosses, and the respective figure number(s). 
Corresponding data can be found in a dataset on the Data Repository of the University of Stuttgart (DaRUS) \citep{DarusData2023}.} \label{Fig:RegimemapLambda}
\end{figure*}

In order to test the above first order criteria, the outcomes produced by all the experimentally and numerically investigated collisions are classified according to the three possible types of liquid distributions. %
Then, for each collision, $\lambda$, $\Lambda$ and $N$ values are computed before being plotted in the form of two dimensional maps using either ($\lambda$; $N$) or ($\Lambda$; $N$), as shown by Figs.~\ref{Fig:RegimemaplambdaNoRho} and \ref{Fig:RegimemapLambda}, respectively. %
The experimental data are represented by full symbols while numerical ones are shown by empty symbols. %
The error bars, depicted for both experimental and numerical data, represent the uncertainties on the relative velocity at the head-on separation threshold. %
The lower bound is provided by the highest velocity leading to encapsulation and the upper bound by the lowest velocity resulting in separation. %
Those bounds influence the parameters $\lambda$ and $\Lambda$ via the value of the relative velocity taken to evaluate the Weber number, $We_i$. %
The black crosses correspond to the illustrative images presented in this work and are plotted together with their respective figure number. %
They are systematically found on the left side of the horizontal error bars since they all illustrate the first head-on separation. %
Experimental uncertainties in the measurements of the liquid properties whose values enter in the evaluation of both parameters are not represented here. %
Crossing separation is represented by red squares, single reflex separation by green diamonds and reflexive separation by blue triangles. %
Furthermore, cases of partial wetting ($S<0$) are indicated by black circles. %

Looking at the resulting maps evidences the efficiency of the proposed criteria. %
Indeed, even without the bending correction applied to $\lambda$, cf.\ Fig.~\ref{Fig:RegimemaplambdaNoRho}, three  separated domains, that correspond to the three possible types of liquid distributions, can be roughly distinguished. %
As expected, crossing separation is observed when $\lambda$ is large, i.e., when the inertial progression of the outer liquid in the axial direction normalized by its radial spreading is large. %
This observation agrees very well with the interpretation based on the importance of the flow of the outer liquid around the inner one and the distribution it reaches at the end of the first collision phase. %
This is especially true if the relative axial displacements of the outer liquid were corrected for the potential bending of the disk, i.e., when $\Lambda$ is used instead of $\lambda$, cf.\ subfigure b). %
It also suggests that the bending of the disk, observed in Fig.~\ref{Fig:RhoRatio}, must be accounted for as the combination of $\lambda$ and $N$ fails in distinguishing crossing separations from single reflex and reflexive separation in every case. %
Especially one point corresponding to $\rho_i/\rho_o=0.625$, which clearly results in single reflex separation is found in the domain where crossing is expected. Applying the correction for the bending to this numerical data point decreases the estimation from  $\lambda=0.25$ to $\Lambda=0.16$ and suppresses this discrepancy. %
More generally, the correction of $\lambda$ into $\Lambda$ sharpens the limit between crossing separation and the other two regimes and justifies, a posteriori, our approach. %
The transition is found for $\Lambda \approx 0.2$, and the other parameter, $N$, has only second order effects. %
For a given $\Lambda$ close to $0.2$, the crossing regime is favored by decreasing values of $N$. %
This is in agreement with our interpretation, which attributes small values of $N$ to rather slow and independent recoil of the outer liquid in comparison to the inner droplet. %
As illustrated in Fig.~\ref{Fig:SimExampleRelaxation}(a) and (c), the uncorrelated recoil of the outer and inner phase favors the pinch-off of the outer liquid from the compound and leads respectively to crossing and reflexive separations. %

For small values of $\Lambda$, typically below $0.2$, only single reflex and reflexive separation are found. %
The distinction between these two types of separation requires further comparison of the associated values of $N$. %
In the case of large $N$, i.e., when it is expected that both the inner and outer phases follows similar recoil, single reflex is observed. %
In contrast, for smaller $N$, typically  below $\approx 0.4$, reflexive separation is observed in agreement with our interpretation based on the decoupling of the relaxation of both phases.

Interestingly, the fact that partial wetting is employed instead of total one appears not to have a strong influence of the collision outcomes, at least in the studied range of spreading parameters. 

\section{Conclusion} \label{Sec:Conclusion}
The liquid distribution after head-on separation of two colliding immiscible droplets in air was investigated. %
Depending on the composition of the daughter droplets, three head-on separation mechanisms can be distinguished, namely crossing, single reflex or reflexive separation. %
While two compound droplets result from a single reflex separation, crossing and reflexive separation result in one compound droplet and one pure droplet of the encapsulating liquid with lower surface tension. %
This droplet is found on the impact side of the outer liquid droplet in case of reflexive separation, and on the opposite side for crossing separation. %
The paper aims at predicting the separation mechanism based on the liquid properties and collision parameters only. \newline
Using complementary experimental investigations and numerical simulations, a comprehensive data set was generated, covering all three mechanisms and wide parameter ranges with Weber numbers from $2.6$ to $720$. %
The Reynolds numbers range from $66$ to $1118$, and the spreading parameter from $-5.7$ to $43.6~{\mathrm{mN/m}}$, thus representing partial and total wetting liquid combinations. %
\\

Our analysis revealed that the liquid distribution in the case of crossing separation differs from the other two mechanisms already at the end of the first phase of the droplet collision (compression phase). %
Crossing can only occur if the encapsulating liquid flowed sufficiently far around the inner droplet once the collision complex reached its maximum diameter, $D_\mathrm{max}$. %
Different densities lead to a bending of the inner liquid in the lamella, which must also be considered, as it promotes (convex) or hinders (concave) the progression of the encapsulating liquid. %
Thus, based on a comparison of the corresponding length scales, we derived the first dimensionless parameter, $\Lambda= \frac{\rho_i}{\rho_o}\frac{1}{\sqrt{We_i}}$, to distinguish single reflex and reflexive separation from crossing separation. %
The transition is found for $\Lambda \approx 0.2$. \\
The distinction between single reflex and reflexive separation, on the other hand, is made during the subsequent relaxation phase. %
While in case of single reflex separation the already encapsulated inner droplet neckles, in the case of reflexive separation only the encapsulating fluid breaks up, which is attributed to the interplay of viscous tangential stresses and capillary pressure jumps. %
This is reflected by the second dimensionless parameter, $N={\nu_o/\nu_i}~{\sigma_o/\sigma_{io}}$. %
Values of $N$ typically above $0.4$ represent single reflex separation. %
Interestingly, the wetting behaviour, represented by the spreading parameter $S$, does not appear to influence the collision outcome significantly.

The derived criteria are tested by classifying all experimentally and numerically investigated collisions. %
For this purpose, the values for $\Lambda$ and $N$ computed from these data were plotted in form of a regime map. %
It is shown that the proposed dimensionless parameters have a predictive value for distinguishing the three head-on separation mechanisms from each other for the manifold liquid combinations tested.

\FloatBarrier

\begin{acknowledgments}
The authors gratefully acknowledge the funding by Deutsche Forschungsgemeinschaft (DFG, German Research Foundation) under \href{https://www.simtech.uni-stuttgart.de/exc/research/pn/pn1/}{Germany’s Excellence Strategy - EXC 2075 – 390740016}. 
The simulations were conducted on the supercomputer \href{https://www.hlrs.de/solutions/systems/hpe-apollo-hawk}{HPE Apollo (Hawk)} at the \href{https://www.hlrs.de}{High-Performance Computing Center Stuttgart (HLRS)} under the grant no.~FS3D/11142. The authors kindly acknowledge the granted ressources and support.
\\
The authors gratefully acknowledge the funding of the SFB-TRR75 (project number 84292822) by the DFG for having supported the Summer Program 2019, which is at the origin of this work.  
We want to thank Dr. David Baumgartner who provided illustrative images of G50-SOM5 and G50-SOM20 collisions.
\end{acknowledgments}

\section*{Data Availability Statement}
The data of the simulation setups and tabulated data to recalculate the regime map data are published as a data set on the Data Repository of the University of Stuttgart (DaRUS) under https://doi.org/10.18419/darus-3594 \citep{DarusData2023}. Further data is available on request.

\appendix

\bibliography{literature}

\end{document}